\documentclass[a4paper,fleqn]{cas-sc}

\usepackage[authoryear]{natbib}

\usepackage[utf8]{inputenc}
\usepackage{caption}
\usepackage{subcaption}
\usepackage{bm}

\begin{document}
\let\WriteBookmarks\relax
\def\floatpagepagefraction{1}
\def\textpagefraction{.001}
\shorttitle{Optimization-Based Path-Planning for Connected and non-Connected Automated Vehicles}
\shortauthors{P. Typaldos et~al.}

\title [mode = title]{Optimization-Based Path-Planning for Connected and non-Connected Automated Vehicles}

\author[1]{Panagiotis Typaldos}
\cormark[1]

\author[1]{Markos Papageorgiou}
\author[1]{Ioannis Papamichail}

\address[1]{Dynamic Systems and Simulation Laboratory, School of Production Engineering \& Management, Technical University of Crete, 73100, Chania, Greece}

\cortext[cor1]{Corresponding author}

\begin{abstract}
A path-planning algorithm for connected and non-connected automated road vehicles on multilane motorways is derived from the opportune formulation of an optimal control problem. In this framework, the objective function to be minimized contains appropriate respective terms to reflect: the goals of vehicle advancement; passenger comfort; and avoidance of collisions with other vehicles, of road departures and of negative speeds. Connectivity implies that connected vehicles are able to exchange with each other (V2V) or the infrastructure (V2I), real-time information about their last generated path. For the numerical solution of the optimal control problem, an efficient feasible direction algorithm is used. To ensure high-quality local minima, a simplified Dynamic Programming algorithm is also conceived to deliver the initial guess trajectory for the feasible direction algorithm. Thanks to low computation times, the approach is readily executable within a model predictive control (MPC) framework. The proposed MPC-based approach is embedded within the Aimsun microsimulation platform, which enables the evaluation of a plethora of realistic vehicle driving and advancement scenarios. Results obtained on a multilane motorway stretch indicate higher efficiency of the optimally controlled vehicles in driving closer to their desired speed, compared to ordinary Aimsun vehicles. Increased penetration rates of automated vehicles are found to increase the efficiency of the overall traffic flow, benefiting manual vehicles as well. Moreover, connected controlled vehicles appear to be more efficient compared to the corresponding non-connected controlled vehicles, due to the improved real-time information and short-term prediction.

\end{abstract}




\maketitle

\section{Introduction}
In the past decade, automated driving has attracted strong interest in industry and scientific community. This is fostered by strong technological advancements, compared to which human driving capabilities appear limited in terms of perception of the driving environment, reaction time, and real-time decision efficiency. In addition, variations in driving behavior from person to person or short inattention at high speeds may result in accidents. In fact, the vast majority of road accidents are attributed to human error on the account of, e.g., insufficient sensory information, lack of attention, shortcomings in driving skill or reckless driving. Each year, road accidents result in approximately 1.35 million fatalities and leave some 50 million of injured or disabled worldwide. Road congestion is another major issue, causing excessive delays, fuel consumption and emissions around the globe \citep{goniewicz2016road, montanaro2019towards}. On the other hand, vehicle automation is a challenging area due to the variety and complexity of real-world environments, including avoidance of static and moving obstacles, compliance with traffic rules and consideration of human driving behavior aspects \citep{gu2014toward}.

With the recent advances in vehicle communications, either vehicle-to-vehicle (V2V) or infrastructure-to-vehicle (I2V), new communication channels and a wide range of information becomes available in real time. Connected automated vehicles (CAVs) may receive or exchange relevant information, including their state, the current traffic conditions, the next switching time of a traffic signals etc. This extended information entails better knowledge of the driving conditions for CAVs and may be beneficial in various driving situations concerning road safety, flow efficiency and environmental sustainability \citep{sjoberg2017cooperative, tian2018performance}. Specifically, automated driving may improve various driving aspects, such as lane-changing, obstacle avoidance, forming of platoons with short inter-vehicle distances and the application of smoother acceleration or deceleration.

The development of fully automated driving algorithms is inherently related to planning and updating a vehicle path, which should be efficient, collision-free and user-acceptable. Planning such a path can be seen as a trajectory generation problem, i.e., creation in real time of a quasi-continuous sequence of states that must be tracked by the vehicle via appropriate steering, throttle and braking actions. Inspired by earlier studies on motion planning of robot vehicles in other contexts (e.g. \citep{gonzalez2015review}) and driven by rapid implementations and optimistic forecasts of CAV technologies (e.g. \citep{nagy2001trajectory, gu2002neural}), studies on CAV trajectory optimization in the road traffic context have boomed in the past decade. However, path-planning for road vehicles is a difficult task, since speeds are very high, and safety of the passengers must be guaranteed. Thus, automation on roads calls for sophisticated approaches, such as optimal control methods, advanced feedback control or reinforcement learning \citep{claussmann2019review, haydari2020survey}.

In the related literature, many works exist for trajectory generation of longitudinal motion for CAVs, considering cooperative adaptive cruise control (CACC) and platooning \citep{dey2015review, wang2018review}; cooperative merging at highway on-ramps \citep{rios2016survey, ntousakis2016optimal}; speed harmonization on highways \citep{ghiasi2017speed, malikopoulos2018optimal} etc. On the other hand, there are relatively few studies on CAV trajectory optimization considering lane-changing \citep{wan2018review}, which indicates that optimization of CAV trajectories in both longitudinal and lateral directions is worth additional investigations. Some of these studies approach the problem with optimal control and model predictive control (MPC), while avoidance of collision with other vehicles is handled through potential-field like functions \citep{dixit2019trajectory, rasekhipour2016potential, wang2018local, jalalmaab2015model, makantasis2018motorway}. However, these works assume limited connectivity, as they only consider the current position and speed of the surrounding vehicles. This means that the future path of the obstacles is considered as a projection of their initial states, e.g. assuming zero acceleration.

This paper explores the impact of vehicle connectivity on the efficiency of advancement of automated vehicles (AVs) and on the efficiency of the emerging traffic flow. The investigations address multi-lane motorways with mixed traffic, comprising both automated and manually driven vehicles at different penetration rates. An optimal control problem is employed for the path-planning of AVs, comprising three main elements:
\begin{itemize}
\item A simple kinematic model describing the vehicle movement process.
\item An objective function to be minimized, which contains respective terms to reflect efficient vehicle advancement; passenger comfort and fuel consumption; avoidance of collisions with other vehicles and of road departures.
\item Short-term prediction of the trajectories of other neighboring vehicles (obstacles), which is crucial for pro-active collision avoidance.
\end{itemize}

For the numerical solution of the optimal control problem (OCP), a very efficient iterative feasible direction algorithm (FDA) is used. To ensure high-quality local minima, a simplified Dynamic Programming algorithm is also employed to deliver the initial guess trajectory for the FDA. Thanks to low computation times, the approach is readily executable within a MPC framework, applying updated initial state and obstacle path prediction at each repetition. MPC has a long history of applications in control, automation, and chemical industries \citep{mayne1988receding, qin2003survey, mayne2014model}. One approach is to solve the OCP at each time step, with a corresponding shift of the planning horizon. An alternative approach is to update path decisions event-based \citep{earl2007decomposition, khazaeni2016event}, which is pursued in the current study.

Vehicle connectivity refers here to the capability of AVs to exchange with each other, in an asynchronous mode, real-time information about their current state (position and speed) and their latest generated path, something that enhances the prediction accuracy for obstacle movement in the short-term future. The MPC-based approach is embedded within the Aimsun micro-simulation platform \citep{Aimsun2019}, which enables driving evaluation in countless appearing driving scenarios. Specifically, we investigate two cases, each of them at different penetrations of AVs:
\begin{itemize}
\item No connectivity: Each AV is aware of the current position and speed of obstacles (via its own sensors). Short-term prediction of obstacle movement is based on extrapolation, assuming zero acceleration.
\item Connected automated vehicles: Each AV is aware of the current position and speed of obstacles; in addition, it receives the path-planning decisions of other automated vehicles, which facilitates the short-term prediction of their movement.
\end{itemize}
In both cases, the short-term movement prediction for manually driven vehicles is based on zero-acceleration extrapolation.

Demonstration results are reported for a motorway pipeline section. The results indicate higher efficiency of the optimally controlled vehicles in driving closer to their desired speed, compared to non-automated vehicles. In addition, increasing penetration rates of controlled vehicles are found to lead to more efficient traffic flow. Specifically, as the penetration rate of controlled vehicles rises, there is a significant increase of the average speed and, consequently, a decrease on the average delay time and travel time for all vehicles. Finally, connected controlled vehicles appear to be more efficient than non-connected controlled vehicles, due to the improved real-time information that enables more pertinent obstacle movement prediction. Also, the emerging traffic flow and even the manually driven vehicles are found to benefit from improved operation of the AVs.

The rest of the paper is organized as follows: Section 2 presents the dynamics of each AV and the components of the objective function that lead to the OCP definition. Section 3, describes the numerical solution algorithms and explains the procedure used for MPC. Section 4 presents the simulation results, while Section 5 concludes this work.

\section{Optimal Control Problem Formulation}
This section describes the proposed path-planning strategy for automated road vehicles on motorways. At first, the simple vehicle kinematic motion dynamics are defined. Then, an objective function is designed, which includes appropriate terms regarding the efficient vehicle advancement, the obstacle and off-road avoidance and the passenger convenience. Finally, the proposed path-planning algorithm is generated based on a combination of FDA and Dynamic Programming techniques.

\subsection{Problem variables and state-equations}

We consider a straight road on a two-dimensional plane, and the vehicle's position on this plane is expressed in global Euclidean coordinates. Each vehicle is described by five state equations, corresponding to the equations of motion, which are expressed in discrete time, assuming time-steps of length $T$, as follows:
 
\begin{align} 
	&x(k+1) = x(k) + v_x(k) \cdot T + \dfrac{1}{2} a_x(k) \cdot T^2 + \dfrac{1}{6} j_x(k) \cdot T^3	\label{eq: state1} \\
	&y(k+1) = y(k) + v_y(k) \cdot T + \dfrac{1}{2} a_y(k) \cdot T^2	\label{eq: state2} \\
	&v_x(k+1) = v_x(k) +  a_x(k) \cdot T + \dfrac{1}{2} j_x(k) \cdot T^2	\label{eq: state3} \\ 
	&v_y(k+1) = v_y(k) + a_y(k) \cdot T	\label{eq: state4} \\
	&a_x(k+1) = a_x(k) + j_x(k) \cdot T	\label{eq: state5}
\end{align}
where $x(k), y(k), v_x(k), v_y(k), a_x(k)$ correspond to the longitudinal and lateral position, the longitudinal and lateral speed and the longitudinal acceleration at time-step $k$, respectively; while the control variables $j_x(k), a_y(k)$ refer to the longitudinal jerk and the lateral acceleration, respectively, which are kept constant for the duration of each time-step $k$; hence the above state equations are derived from the exact time-integration of the corresponding continuous-time differential equations of motion. Note that the consideration of jerk (rather than the acceleration), as a control variable for the longitudinal direction, leads to smoother vehicle trajectories, which consequently improve the convenience of the vehicle's passengers. On the other hand, for the lateral movement, such detail is not necessary, as the lateral speed and movement is only needed for lane changing, and a lane change maneuver is substantially less frequent compared to the continuous longitudinal motion. 

It should be noted that the application of the lateral position in the simulation is in discrete form due to the nature of the micro-simulator used for the evaluation, meaning that, while the proposed approach produces continuous lateral movement, this movement is translated into a change from one lane to another, as will be explained in detail later. Note also that the control variables $j_x(k), a_y(k)$ may be bounded according to vehicle specifications, if necessary.

\subsection{Optimization Objectives}

\subsubsection{Objective Function}
The objective function to be minimized, in the frame of the optimal control problem formulation, includes a number of terms, which consider efficient, convenient and safe driving, each with a weighting parameter to reflect the corresponding priorities. The criterion, which extends over a time horizon of $K$ steps in the future, reads
\begin{equation}{\label{eq: obj. criterion}}
\begin{aligned}
	J = \sum_{k=0}^{K-1} \Big[ w_1 j_x^2(k) +
			w_2 a_x^2(k) +
			w_3 a_y^2(k) +
			w_4 [v_x(k) - v_{d,x}]^2 +
			w_5 [v_y(k) - v_{d,y}]^2 \\
			+	w_6 f_r(y(k)) +
			w_7 \sum_{i=1}^{n} \big[ c_i(x(k), y(k)) \big] +
			w_8 f_{ns}(v_x(k))
	\Big]
\end{aligned}
\end{equation}
where $w_1, \dots, w_8$ are non-negative penalty parameters.

The first three quadratic penalty terms concern the comfort of the passengers, which is related to the magnitude of lateral and longitudinal acceleration, as well as of the longitudinal jerk. Note that the quadratic penalty term of longitudinal acceleration acts also as a good proxy for deriving fuel-minimizing vehicle trajectories \citep{typaldos2020minimization}. More specifically, it has been demonstrated that the simple square-of-acceleration term delivers excellent approximation of fuel-optimal vehicle trajectories, when compared with the use a complex and realistic fuel-consumption model in the objective function. 

The fourth and fifth penalty terms reflect the vehicle advancing goals. These terms account for pre-specified desired, longitudinal and lateral, speeds by penalizing speed deviations from those values. In the current work, $v_{d,x}$ has a positive value, corresponding to the vehicle type or driver choice of desired longitudinal speed; while $v_{d,y}$ is set to zero to minimize unnecessary lateral movements. Note that, $v_{d,y}$ can be set to non-zero values in case, for example, a vehicle is bound to exit the motorway at a downstream off-ramp or in emergency cases, where the controlled vehicles should create appropriate space, e.g., for an ambulance or a fire truck to pass. 

The three last penalty terms, which are related to the lateral road boundaries, the obstacle avoidance and the negative speed bound, respectively, are described in detail in the following.

\subsubsection{Road Boundaries Term}

\begin{figure}
	\centering
		\includegraphics[width=.75\linewidth]{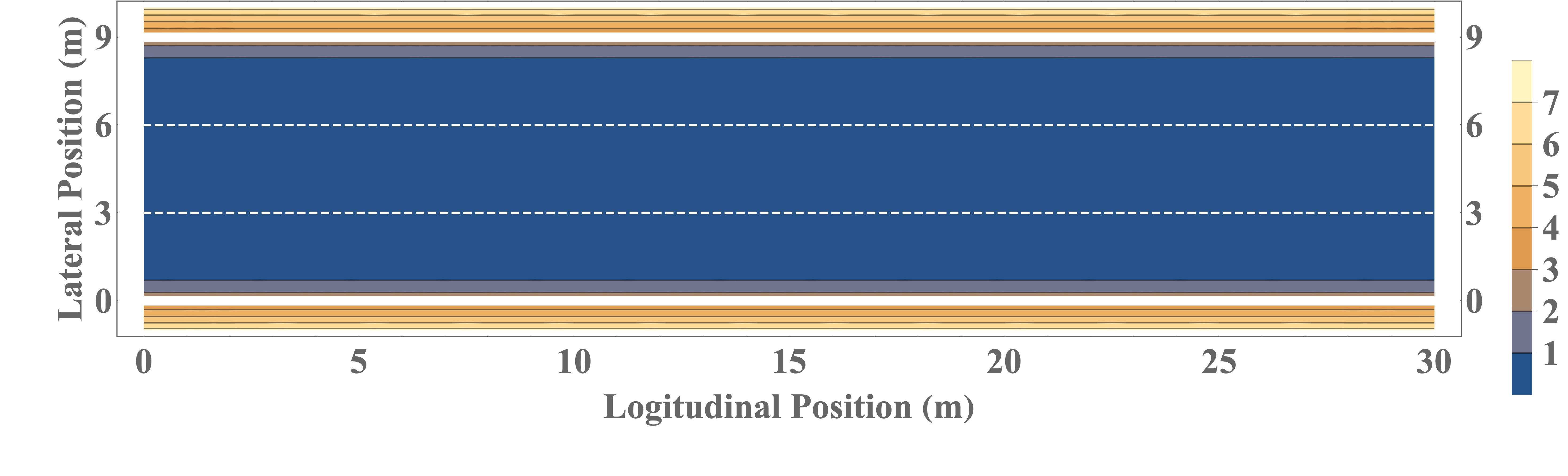}
	\caption{The penalty function of road boundaries. The white solid and dashed lines represent the road and lane boundaries, respectively.}
	\label{FIG:road}
\end{figure}

The penalty function for the road boundaries must be designed so as to disallow the automated vehicle (AV) to depart from the road. To this end, the following smooth quadratic function, dependent on the width $w$ of the road, is adopted, which is equal to zero when the vehicle moves within the road boundaries; while its value increases the more the vehicle would depart from the boundaries

\begin{equation}\label{eq: road-bound}
f_r = \begin{cases}
			 (y - d)^2			&	, y < d \\
			 0						&	, d \leq y \leq w - d \\
			 (y + d - w)^2	&	, y > w - d
		\end{cases}
\end{equation}
where the small positive parameter $d$ is used to prevent the ego vehicle from moving too close to the road boundaries; and $w$ is the road width. Figure \ref{FIG:road} displays the penalty function generated for the road boundaries for a three-lane motorway with $w = 9$ m and $d = 1.5$ m.

\subsubsection{Collision Avoidance Term}

Obstacles correspond, in the motorway case, to other moving vehicles. The penalty function (similar to potential field functions) for obstacle avoidance should feature high values at the gross obstacle space, so that the ego vehicle is repulsed, and potentially unsafe trajectories are suppressed; and low (or virtually vanishing) values outside of that space. To this end, we adopt, for each obstacle, an ellipsoid penalty function, which is best understood in its one-dimensional form $f(x) = (1 + (x / a)^{p_1})^{-1}$. This smooth function has a maximum of 1 for $x = 0$; for $|x|<a$, the function retains values close to 1 (the maximum), while for $|x|>a$, it reduces to very small values or virtually zero. The even integer parameter $p_1$ influences the sharpness of the smooth transition of function values from 1 (for $|x|<a$) and 0 (for $|x|>a$).

In the present case, the function is generalised to two dimensions, and its two respective arguments $\delta_x$ and $\delta_y$ reflect the longitudinal and lateral distances between the ego vehicle and each obstacle $i$, while the respective counterparts of parameter $a$ should be selected based on the vehicle dimensions, taking into account also the vehicle and obstacle speeds for safe car-following distances. Thus, in two dimensions, the following ellipsoid penalty function for each obstacle $i$ is used

\begin{equation}\label{eq:ellipse}
c_i(x,y) = \dfrac{1}{ \big(\frac{\delta_x + s}{ 0.5 \cdot r_x}\big)^{p_1} + \big(\frac{\delta_y}{0.5 \cdot r_y} \big)^{p_2} + 1 }
\end{equation}
where $p_1$ and $p_2$ are positive even integers, $\delta_x=x-x_i$ and $\delta_y=y-y_i$ are the longitudinal and lateral distances from obstacle $i$, $r_x$ and $r_y$ determine the dimensions of the ellipse, and the term $s$ introduces a shift of the ellipse's longitudinal (centre) position, as will be explained in the following.

For safety, the width of the ellipse, $r_y$, should be equal to the width of a motorway lane, $W$, to avoid lateral intrusion of the ego vehicle into a lane occupied by an obstacle. On the other hand, the length of the ellipse, $r_x$, should consider a safe space gap in front and behind of each obstacle $i$. In case the ego vehicle is behind the obstacle, a safe space-gap, equal to $\omega \cdot v_x$, should be maintained between the two vehicles, where $\omega$ is the time-gap parameter \citep{rajamani2011vehicle}. On the other hand, when the ego vehicle is in front of the obstacle, a safe space-gap, equal to $\omega \cdot v_i$, should be considered, with $v_i$ being the longitudinal speed of the obstacle. Moreover, the physical dimensions of both ego and obstacle vehicles, $L_i=(\mu_e+\mu_{o,i})/2$, with $\mu_e$, $\mu_{o,i}$ being the ego and obstacle's $i$ lengths, respectively, should be considered as a minimum safe space-gap in case of zero speeds. Thus, the longitudinal ellipsoid dimension is defined as

\begin{equation}
r_x = \omega \cdot v_x + \omega \cdot v_i + L_i  \label{eq: rx}
\end{equation}

Due to the difference in the longitudinal speeds of the ego vehicle and the obstacle, the above space gaps in front of and behind the obstacle are accordingly different, hence the ellipse is longitudinally asymmetric with respect to the physical centre of the obstacle. Therefore, the ellipse centre must be appropriately shifted, depending on ego vehicle and obstacle speeds difference. Specifically, the position of the ellipse centre is expressed as the summation of $\delta_x$ and the shift term $s$ given by

\begin{equation}
s = \dfrac{\omega \cdot (v_x - v_i)}{2}
\end{equation}

\begin{figure}
	\centering
	\begin{subfigure}[b]{0.6\textwidth}
		\centering
		\includegraphics[width=\textwidth]{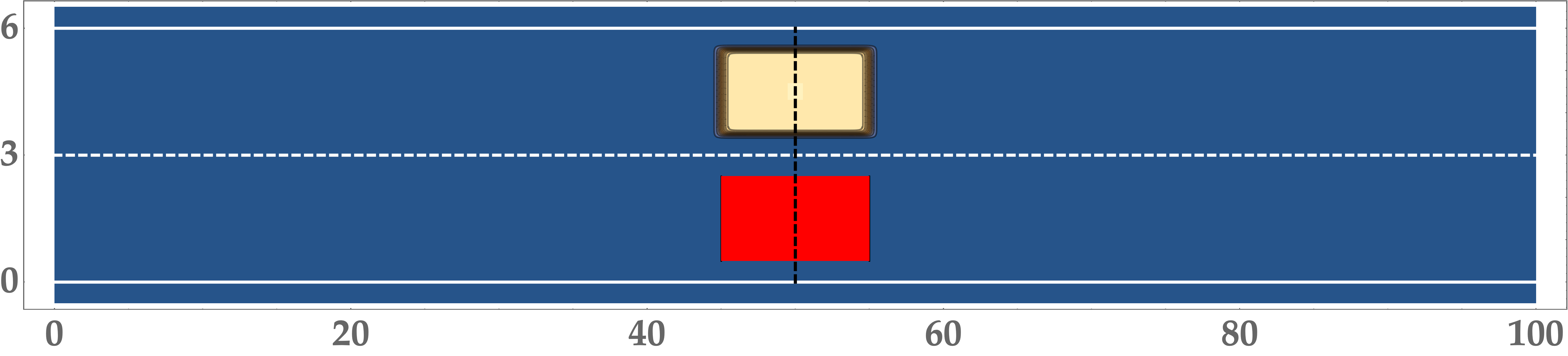}
		\caption{$\omega = 0$ s or ($v_x = 0$ m/s and $v_i = 0$ m/s)}
		\label{fig: ellipse-1}
	\end{subfigure}	
	\begin{subfigure}[b]{0.45\textwidth}
		\centering
		\includegraphics[width=\textwidth]{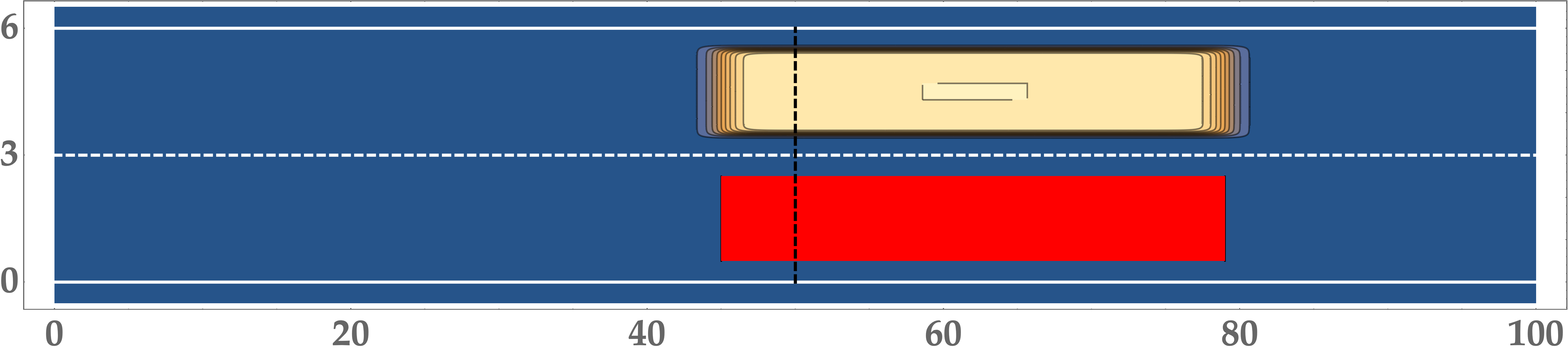}
		\caption{$\omega = 1.2$ s, $v_x = 0$ m/s and $v_i = 20$ m/s}
		\label{fig: ellipse-2}
	\end{subfigure}	
	\begin{subfigure}[b]{0.45\textwidth}
		\includegraphics[width=\textwidth]{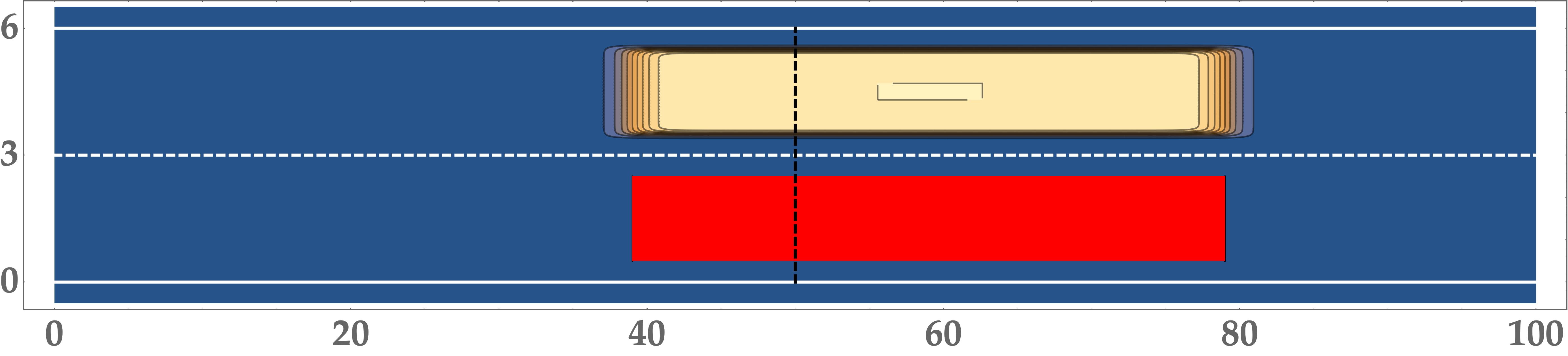}
		\caption{$\omega = 1.2$ s, $v_x = 5$ m/s and $v_i = 20$ m/s}
		\label{fig: ellipse-3}
	\end{subfigure}
	\begin{subfigure}[b]{0.45\textwidth}
		\centering
		\includegraphics[width=\textwidth]{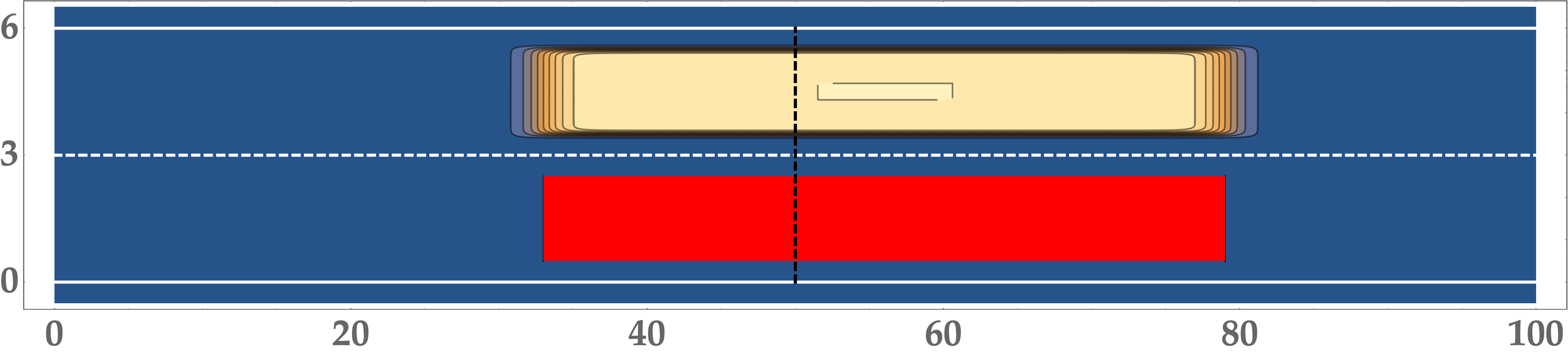}
		\caption{$\omega = 1.2$ s, $v_x = 10$ m/s and $v_i = 20$ m/s}
		\label{fig: ellipse-4}
	\end{subfigure}	
	\begin{subfigure}[b]{0.45\textwidth}
		\includegraphics[width=\textwidth]{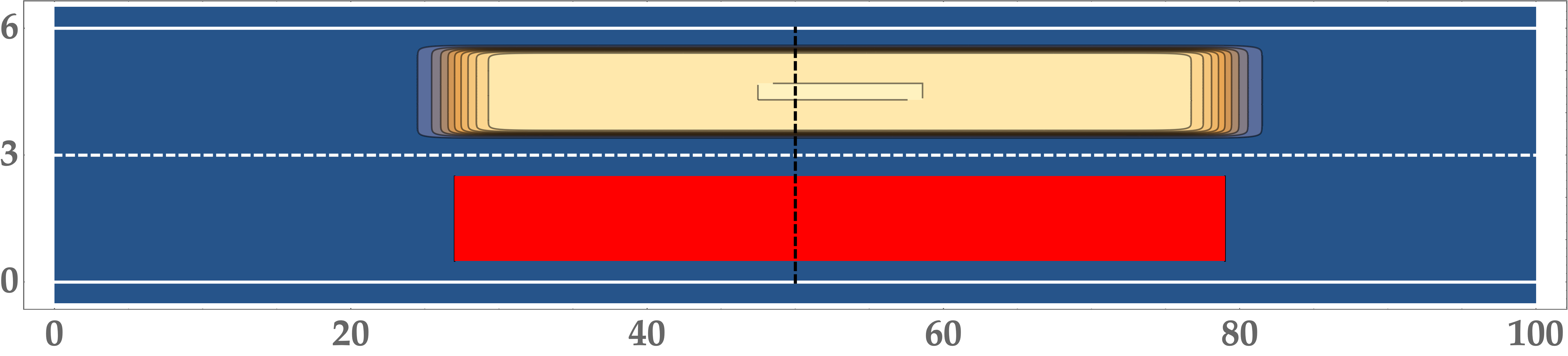}
		\caption{$\omega = 1.2$ s, $v_x = 15$ m/s and $v_i = 20$ m/s}
		\label{fig: ellipse-5}
	\end{subfigure}
	\begin{subfigure}[b]{0.45\textwidth}
		\centering
		\includegraphics[width=\textwidth]{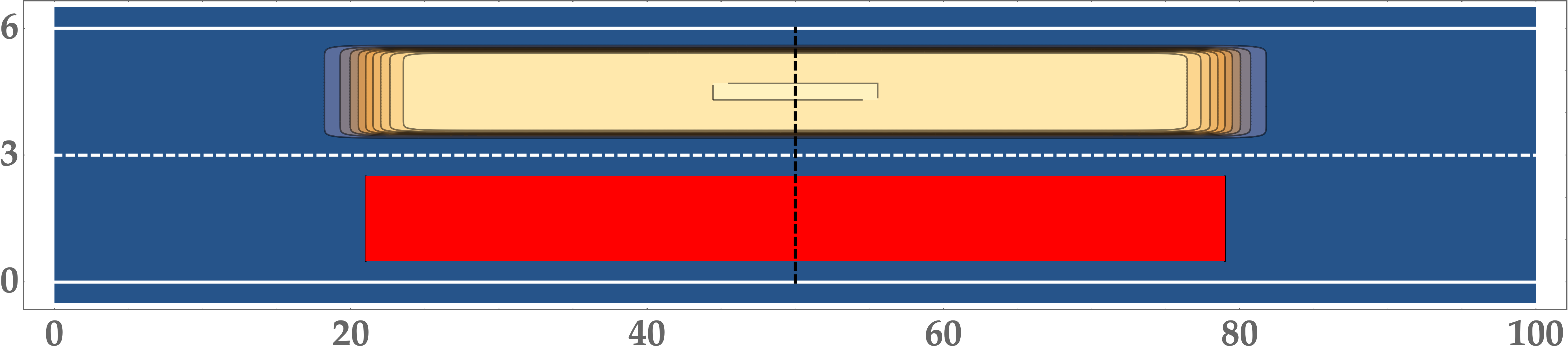}
		\caption{$\omega = 1.2$ s, $v_x = 20$ m/s and $v_i = 20$ m/s}
		\label{fig: ellipse-6}
	\end{subfigure}	
	\begin{subfigure}[b]{0.45\textwidth}
		\includegraphics[width=\textwidth]{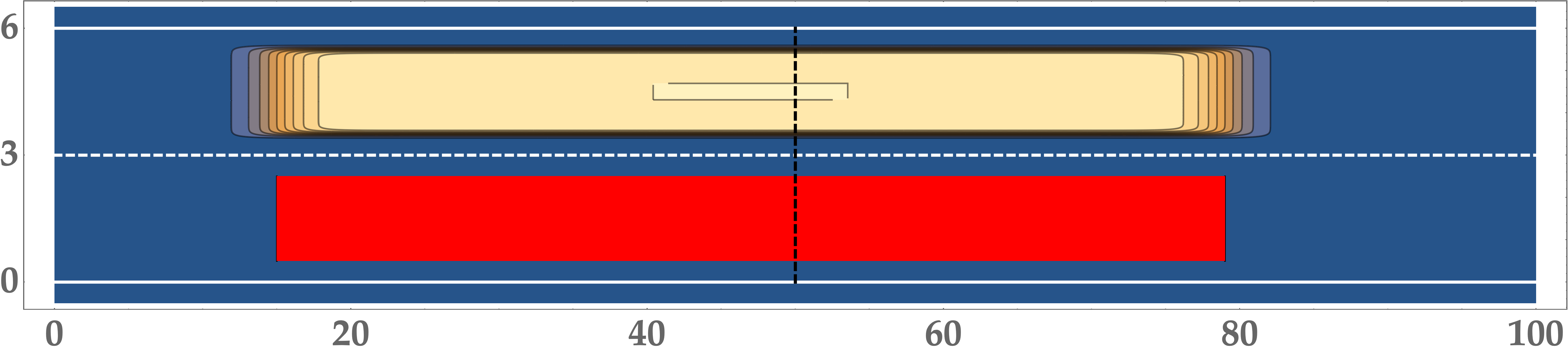}
		\caption{$\omega = 1.2$ s, $v_x = 25$ m/s and $v_i = 20$ m/s}
		\label{fig: ellipse-7}
	\end{subfigure}

	\caption{Illustration of the collision avoidance term (orange ellipsoid), $c_i (x,y)$, for different ego vehicle speeds with time gap $\omega = 1.2$ s and obstacle speed $v_i= 20$ m/s, compared to the corresponding space gap of the constant time gap policy (red rectangle). The lateral dashed line indicates the physical vehicle centre location.}
	\label{FIG: ellipse-phases}
\end{figure}

In Figures \ref{fig: ellipse-1}-\ref{fig: ellipse-7}, different cases for the proposed ellipsoid are displayed for illustration. We assume an obstacle with its centre point at 50 m. The ellipsoids in each case are displayed in orange color and are contrasted to the corresponding space gaps resulting from the constant time-gap policy (red rectangle) \citep{rajamani2011vehicle}. In Figure \ref{fig: ellipse-1}, both the ego vehicle and the obstacle have zero speed, leading to a symmetric ellipse shape around the obstacle centre, as both the front and the back part around the vehicle centre (dashed black line) are equal to $L_i$. Note that the shape of the ellipse is close to a rectangle due to the use of high values for both exponents $p_1$ and $p_2$ in (8) (e.g. $p_1=p_2=18$). In Figures \ref{fig: ellipse-2}-\ref{fig: ellipse-7}, using a time-gap value $\omega = 1.2$ s and a longitudinal obstacle speed $v_i$  = 20 m/s, it is observed how the ellipse shape changes, and the ellipse centre shifts in dependence of the ego vehicle speed. Specifically, when the ego vehicle has zero speed (\ref{fig: ellipse-2}), the centre of the ellipse is shifted to upstream to accommodate a back space-gap equal to the average length of both vehicles; while the front part of the ellipse equals the space-gap, which depends on the speed of the obstacle. For higher ego vehicle speeds, it can be observed that the length of the ellipse increases, and, at the same time, its centre shifts downstream. In the specific representation, the length of the front part of the ellipse does not change, as the obstacle speed remains constant.

Potential obstacles, to be considered in the sum of the collision avoidance term in \eqref{eq: obj. criterion}, are all vehicles around the ego vehicle that might appear on its way, causing potentially a crash risk, during the considered planning horizon. Thus, all vehicles within a longitudinal zone in front and behind the ego vehicle should be considered as potential obstacles, and the length of this zone is taken proportional to the ego vehicle's desired longitudinal speed times the time horizon.   

\subsubsection{Negative Speed Bound Term}
\begin{figure}
	\centering
		\includegraphics[width=.65\linewidth]{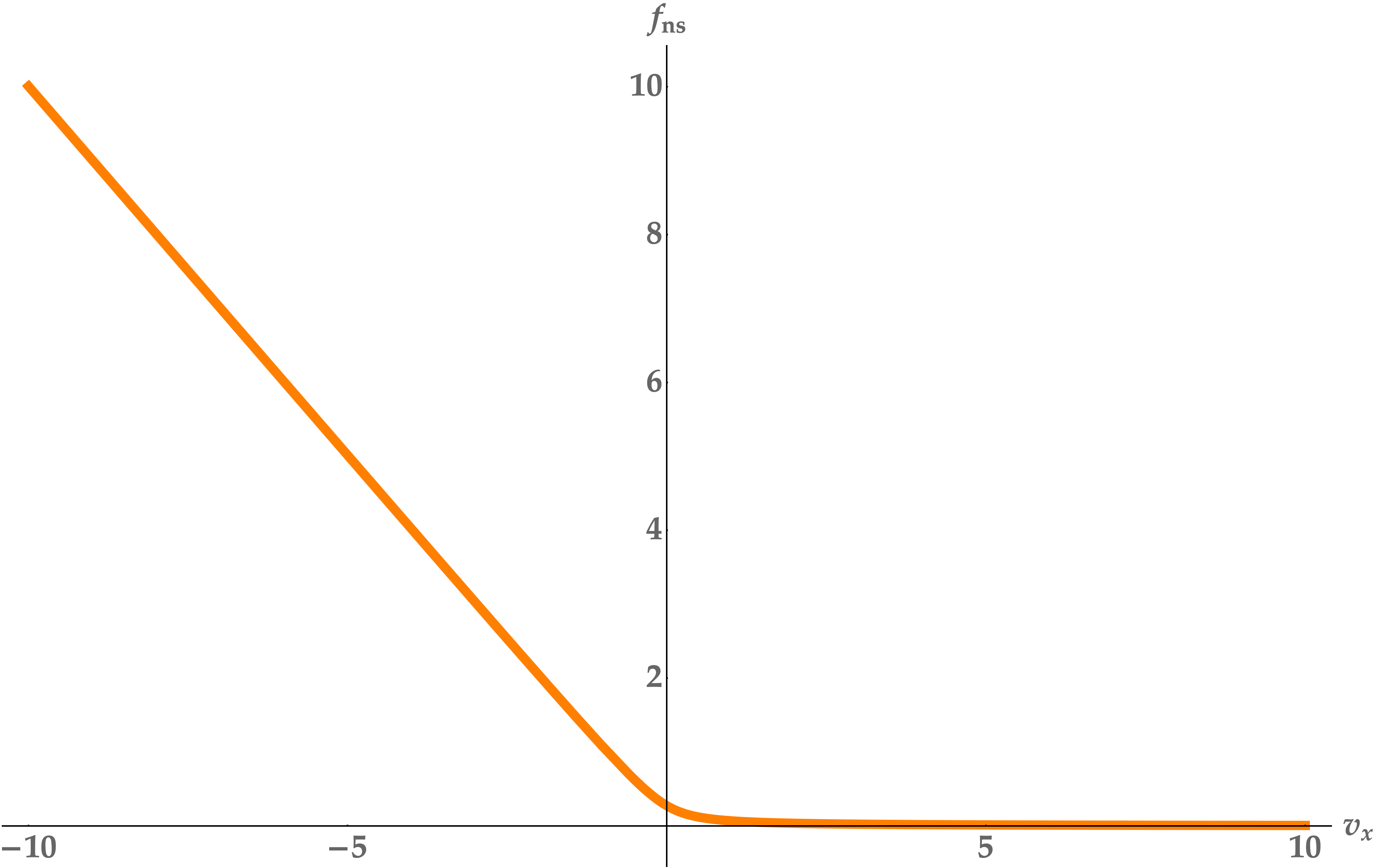}
	\caption{Smooth penalty function for avoidance of negative longitudinal speeds with smoothing parameter $\epsilon =0.1$.}
	\label{FIG:neg-speed}
\end{figure}

The explicit consideration of state constraints in the optimal control problem is challenging for the numerical solution algorithm. Therefore, the negative-speed penalty term is introduced in the objective criterion to penalize negative longitudinal speed values. To this end, a smooth function $g(a,b)$ that features high values when $a$ is negative and turns to $b$ for positive values of $a$ (see Figure \ref{FIG:neg-speed}), is considered. This function reads as follows:

\begin{equation}{\label{eq: neg-speed}}
g(a) = \dfrac{1}{2} (-a-b+\sqrt{(b-a)^2+\epsilon})
\end{equation}
with $\epsilon$ being a smoothing parameter. Based on \eqref{eq: neg-speed}, the negative speed penalty term, $f_{ns}$, is expressed as:
\begin{equation}
f_{ns}(v_x(k)) = ( -v_x(k) + \sqrt{(-v_x(k))^2 + \epsilon} )
\end{equation}

\subsection{Problem Formulation}
In conclusion, the path-planning problem may be formulated as an optimal control problem. The difference equations \eqref{eq: state1}-\eqref{eq: state5} may be organised in the following vector form
\begin{equation}{\label{eq: states}}
	\bm{x}(k+1) = \bm{f}[ \bm{x}(k), \bm{u}(k), k ], \qquad k=0,\dots, K-1
\end{equation}
where $\bm{x}$ and $\bm{u}$ are the system states and control variable vectors, respectively. With known initial state $\bm{x}(0) = \bm{x}_0$, the optimal control problem consists in minimizing the objective function \eqref{eq: obj. criterion} subject to \eqref{eq: states} and control bounds

\begin{equation*}
\bm{u}_{\min}(k) \leq \bm{u}(k) \leq \bm{u}_{\max}(k), \qquad k=0,\dots, K-1
\end{equation*}
where $\bm{u}_{\min}, \bm{u}_{\max}$ are the constant lower and upper control bounds.
Although expressing a dynamical process, the above minimization problem is, from a mathematical point of view, a Nonlinear Programming Problem (NLP) due to the discrete-time nature of the involved process model, see (\citep{papageorgiou2016feasible}, \citep{papageorgiou2015optimierung}). However, thanks to the structure of the state equation \eqref{eq: states}, which allows for the state variables to be efficiently eliminated as functions of the control variables, the optimisation problem may be solved, by use of reduced gradients, in the reduced space of the control variables much more efficiently, compared to a general NLP problem with equal dimension.

\section{Numerical Solution and Model Predictive Control}

\subsection{Feasible Direction Algorithm (FDA)}
The solution of the formulated OCP is computed by use of the very efficient feasible direction algorithm (FDA) \citep{papageorgiou2016feasible, typaldos2020minimization}, which exploits the structure of the state equations to map the OCP into an NLP problem in the reduced space of control variables. Thus, the algorithm attempts the calculation of a control trajectory $\bm{u}(k), k = 0,\dots,K-1$, which corresponds to a local minimum of the cost function, in the $mK$-dimensional space, where $m$ is the number of control variables. This marks a substantial reduction of the problem dimension, as the state variables are eliminated.

The algorithm is iterative, starting with an initial-guess feasible control trajectory; feasible meaning that it satisfies all state equations and control inequality constraints. At each iteration, using reduced gradient information, a descent search direction in the (reduced) $mK$-dimensional control space is calculated based on conjugate gradients (or quasi-Newton methods). Subsequently, a line-search procedure delivers the optimal step along the search direction, and this optimal step leads to an enhanced feasible control trajectory, with improved objective function value. This enhanced trajectory is fed to the next iteration; and so forth, until a sufficiently low reduced-gradient norm is obtained, which marks convergence to a virtually optimal control trajectory. The algorithm guarantees improved objective values at each iteration and features global convergence, from any starting control trajectory, to a local minimum. 

\subsection{Dynamic Programming (DP)}
The optimal trajectory produced by the FDA may correspond to a local minimum, which depends on the utilized initial guess trajectory. In particular, if the initial guess trajectory does not feature a lane change, then, in most cases, the trajectory resulting from the FDA iterations keeps the vehicle on the same lane, optimising its car-following behavior only (local minimum), although a lane change at an appropriate time within the considered time horizon might in some cases decrease the value of the objective function. In fact, when trapped in such local minima, the algorithm has no local gradient-based incentive to explore the possibility of a lane change, as this does not appear beneficial locally due to the shape of the ellipsoid function of the collision avoidance term \eqref{eq:ellipse}, in combination with the penalisation of lateral speed and acceleration, leading to trajectories with no lateral movement. 

In order to overcome this difficulty, Dynamic Programming (DP) is used to produce the initial guess trajectory. The DP methodology is known to deliver a globally optimal solution trajectory for optimal control problems. However, DP is characterized by computational cost, which increases exponentially with the problem dimensions; and this cost is indeed too high in the present application for efficient real-time path-planning. Therefore, we employ DP to solve, with very low computation times, a simplified version of the path-planning optimal control problem. The produced rough, but globally optimal simplified-problem solution is then used as an initial guess trajectory to be refined by the FDA. In particular, such rough, but globally optimal solutions of a simplified problem were found to include lane changes in driving situations where a lane change is indeed beneficial.

A first simplification in the DP problem concerns a larger step size $T$ (1 s) versus the four times smaller time step (0.25 s) used in FDA and deemed appropriate for path-planning of AVs. As a consequence, the number $K$ of time steps within each planning horizon decreases accordingly. In addition, the state equation \eqref{eq: state5} is dropped, and the longitudinal acceleration $a_x$ is used as a control variable instead of the jerk. The lateral vehicle motion is also simplified by dropping state equations \eqref{eq: state2} and \eqref{eq: state4} and assuming that the vehicle's lateral position is discrete and lane-based. Specifically, the lateral control is limited to three distinct values at each time step, namely $[-1,0,1]$, meaning that the vehicle can only apply a lane change towards the adjacent left or right lanes or stay on the same lane. In addition, only one lane change is allowed at each planning horizon, in order to reduce the amount of options to be explored by DP. Finally, the longitudinal acceleration is also roughly discretised and may obtain one out of three values, namely $[-3,0,3]$ m/s$^2$.

Given the above modifications and the fact that state constraints may be directly handled by the DP algorithm, the cost function used with DP is simpler than the corresponding cost function of the OCP and is described as follows

\begin{equation}{\label{dp cost}}
	J_{DP} = \sum_{k=0}^{K-1} [ w_1 a_x^2(k) + w_2 u_y^2(k) + w_3 (v_x(k) - v_d)^2 ]
\end{equation}

where $a_x$ and $u_y$ are the control variables, corresponding to the longitudinal acceleration and the discrete lateral movement, which reflects the lane changes. Collision avoidance, road departure avoidance and suppression of negative speeds are taken care by the DP algorithm directly, as will be discussed later.

Due to the simplifications introduced for both control variables (using longitudinal acceleration and lane changing, instead of longitudinal jerk and lateral acceleration), the rough discretisation and the decrease of the number of time steps, the resulting DP-optimal control trajectories have to be processed appropriately, in order to be consistent with FDA control inputs. To this end, DP's control decisions, including the lateral control values, are mapped into the corresponding values in the continuous two-dimensional space; and Euler forward integration is used to obtain the corresponding initial guess trajectory for the FDA.

After discretisation of the longitudinal position, speed and acceleration with consistent respective increments, the standard DP algorithm is designed as follows. The algorithm starts with time step $K-1$ and advances backward, step-by-step. At every step $k$, all discrete states are branched into all possible transitions (reflecting all combinations of the discrete $a_x$ and $u_y$ values), whereby infeasible transitions (road departure, obstacle collision, negative longitudinal speed) are ignored. For each feasible discrete state, the corresponding optimal controls, along with the corresponding optimal cost-to-go value, are stored. The algorithm ends, when the initial state at $k=0$ has been evaluated. Eventually, starting from the given initial state and progressing forward by following the respective optimal controls at each encountered discrete state, the optimal trajectories are obtained.

In view of the positive semi-definite nature of the objective function \eqref{dp cost}, an alternative, forward-branching procedure, employing branch-and-bound methods, was also tested for the solution of the simplified DP problem. In this approach, the initial state is first branched via all possible transitions (control combinations) to reach corresponding states at the next time step $k=1$, whereby transitions to infeasible states are ignored. The states obtained from branching are evaluated, regarding the cost to reach them from the initial state. The algorithm continues, and, each time, the most efficient (lowest cost from initial state) open (not yet branched) state (of any time step) is branched next. The new branched states are evaluated, and so forth, until the end of the time horizon has been reached with a cost that is lower than the cost of any other open state of any time $k$. Remarkably, this branch-and-bound procedure was found to lead to identical solutions as the standard DP algorithm, albeit within a computation time ten times lower, on average, compared to the DP solution. Specific run times for all employed algorithms are provided in the next section.

\subsection{Safety Override}
The proposed optimal control approach does not guarantee crash-free vehicle trajectories. Thus, under circumstances, a path that includes a crash may be produced. This is mainly due to the conflict of the collision avoidance term versus the advancing term, reflecting the need to drive at a pre-specified desired speed. These two terms may be conflicting, as the first term may be striving to decrease the ego vehicle speed, in presence of a slower leading obstacle, while the second term is striving to increase the speed towards the desired speed. A balance between these two terms is typically reached (through the corresponding penalty weights in \eqref{eq: obj. criterion}), which guarantees efficient vehicle advancement while suppressing collisions. However, under extreme conditions, e.g. in high density scenarios, the possibility of a collision cannot be utterly excluded. 
In order to avoid such decisions and ensure safety, an emergency rule is activated if the solution procedure produces an unsafe path. In this case, the just generated path is dropped, and a new one is generated, with reduced desired speed and planning horizon. Specifically, the desired speed is set equal to a percentage of the leading obstacle's speed (e.g. 95\% of obstacle's current speed) and the planning horizon is reduced by half. Both these measures reduce the size of the desired-speed term and enable crash-free vehicle advancement. In particular, the reduction of the planning horizon is helpful because the objective function \eqref{eq: obj. criterion} is additive over the time steps; thus, in some rare driving scenarios, it may appear less costly to crash with the leading vehicle for the first few time steps and then achieve all goals for the rest of the planning horizon; instead of avoiding the collision and retaining a high cost each time step due to the desired-speed deviation. Thus, the reduction of both the ego vehicle's desired speed and the planning horizon allows for avoidance of any collision with the obstacles, until the ego vehicle finds sufficient space on an adjacent lane to overtake and approach its desired speed.

\subsection{Model Predictive Control}
Summarising, the presented numerical solution approach requires, as input data, the current (initial) ego vehicle (EV) state, as well as the current and future positions and speeds of obstacle vehicles (OVs); to produce optimal EV controls and states over a future time horizon $KT$. This is an open-loop solution, and, given the dynamic environment (moving OVs), the time horizon should be long enough to anticipate and prepare for future situations and avoid myopic control actions. On the other hand, as time advances, the uncertainties related to the changing environment (actual OV movement) and to the actual vehicle advancement (as compared to the open-loop solution) increase, as increasing deviations from the assumed predictions are inevitable. To address these uncertainties, the open-loop solution procedure may be cast in a model predictive control (MPC) frame, whereby the solution is re-computed online, using the same horizon $KT$ (rolling or receding horizon), whenever substantial changes regarding the initial predictions are detected at any time before the end of the time horizon. The new computation uses updated initial states and updated predictions about the movement of OVs. This calls for computation times smaller than the path update period, something that is indeed satisfied by the presented efficient solution procedure.
Considering traffic flow with many vehicles, as expected in real conditions in the near future and as considered in the subsequent simulation investigations, three types of vehicles are distinguished:

\begin{enumerate}
\item Manually driven vehicles, which, in the simulation investigations, are navigated by the employed microscopic simulator (Aimsun). 
\item Automated vehicles without V2V communication capabilities, which are navigated according to the presented procedure. Such vehicles rely only on their own sensors to sense the current position and speed of other surrounding vehicles of any type; and their paths are predicted simply by assuming that they will keep their current lane and speed fixed over the EV planning horizon $KT$.
\item Connected automated vehicles (with V2V communication capabilities), which are also navigated according to the presented procedure. However, such vehicles may broadcast their latest path decision to other surrounding connected vehicles; and can receive the latest path decisions of those surrounding connected vehicles. Note that this information is broadcasted asynchronously, i.e. a path-planning decision by a vehicle is broadcasted as soon as it is produced. Thus, connected AVs rely also on their own sensors to sense the current position and speed of other surrounding vehicles; but, in addition, they receive the latest path planning decision by other surrounding vehicles of the same type.
\end{enumerate}
	
As mentioned earlier, the EV path is re-generated in real time to address evolving deviations from the last predicted driving conditions. More specifically, a new updated path is generated in the following cases:

\begin{itemize}
\item The EV has driven for the duration of half planning horizon ($KT/2$) according to the last generated trajectory. The plan is then updated, even if no deviations are observed, because application of the second half of the last path may lead to myopic actions.
\item One or more surrounding vehicles deviate substantially from their predicted paths, e.g. an OV changed lane or changed its speed significantly, compared to its predicted movement.
\item A new vehicle enters into the planning zone around the EV, corresponding to a new OV that was not accounted for in the last EV planning.
\item The controlled vehicle cannot track the produced path (e.g. when certain safety-related lane-changing restrictions are violated in the microscopic simulation environment).
\end{itemize}
	
\section{Simulation Testing and Results}

\subsection{Simulation Environment}

\begin{figure}
	\centering
		\includegraphics[width=.75\linewidth]{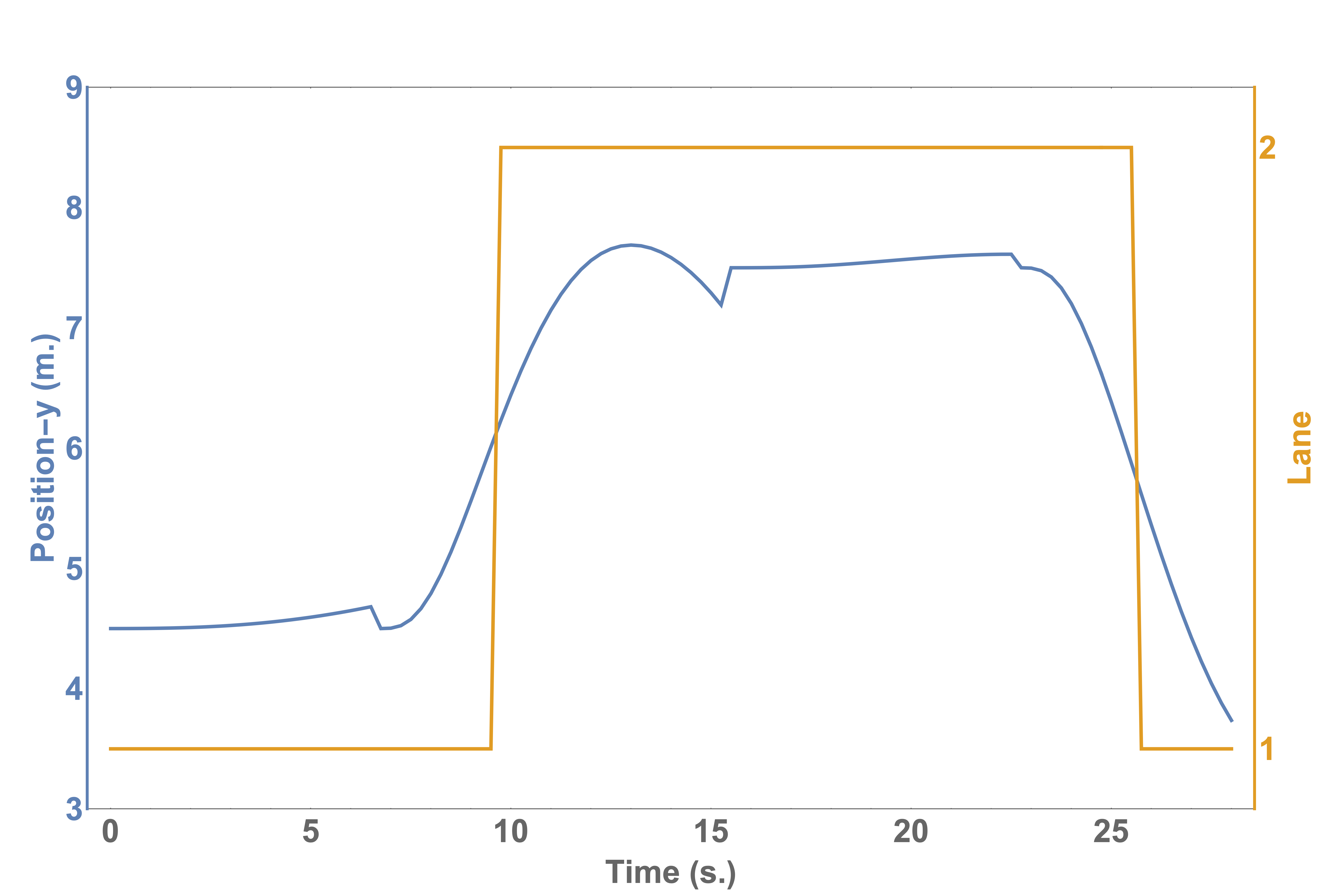}
	\caption{AV lateral position trajectory (blue line) and its mapping into Aimsun lanes (orange lines).}
	\label{FIG:lanes}
\end{figure}

In order to evaluate the proposed path-planning approach, including its MPC-based application, in a realistic environment, the procedure was implemented in Aimsun’s \citep{Aimsun2019} micro-simulation platform with the use of both provided API and SDK tools to integrate the developed path-planning procedures in the traffic context. This implementation enables the investigation of how vehicles, guided by our path-planning approach, interact with each other and with other vehicles, that emulate human driving, in countless driving situations occurring for a variety of traffic conditions.
Due to the discrete nature of the micro-simulator with respect to lateral vehicle movement, the produced path of each AV must be modified appropriately to enable its application within the simulator. Specifically, Aimsun does not allow for continuous lateral vehicle positioning or movement, other than discrete lane assignment and instantaneous (vertical) lane changing. Therefore, the continuous lateral AV movements, produced by the path-planning approach, must be translated appropriately in terms of Aimsun lane positioning. For example, in a three-lane motorway section with each lane being 3 m wide, Aimsun assumes that the right-most lane is lane 1, the middle lane is lane 2 and the left-most lane is lane 3. Thus, the lateral EV position at any time must belong to one of these lanes. On the other hand, in the proposed path-planning approach, lateral vehicle position is continuous (as in real conditions), hence, the AV assignment to a discrete lane in the simulator is effectuated according to the AV lateral position: if the AV lateral position is within a range [0, 3] m, the AV is assigned to the right-most lane; if it is within range [3, 6] m, it is assigned to the middle lane; and for range [6, 9] m, the AV is assigned to the left-most lane. The same applies also in case the AV is executing a lane change, leading to a “vertical” lane change, as required by Aimsun, according to Figure \ref{FIG:lanes}.

For the simulation investigations, two cases were considered, each of them at different penetrations of AVs:

\begin{itemize}
\item No connectivity: Each AV is aware only of the current position and speed of obstacles (via its own sensors).
\item Connected automated vehicles: Each AV is aware of the current position and speed of obstacles; in addition, it receives the path-planning decisions of other AVs, which facilitates more accurate short-term prediction of their movement.
\end{itemize}
	
In both cases, manually driven vehicles are moved according to Aimsun’s Gipps (lane-changing) \citep{gipps1986model} and IDM (car-following) \citep{treiber2013traffic} models. 

All investigations use as a testbed a homogeneous motorway section of 3 km in length, with three lanes (each lane being 3 m wide). Two different levels of inflow, 3.000 veh/h and 5.000 veh/h, into this section are simulated, and vehicle trajectories and traffic conditions are monitored over a simulation horizon of 60 min.  Entering vehicles are randomly assigned their characteristics (type of vehicle, dimensions, desired speed, initial lane and time gap). In particular, vehicle type is selected randomly, according to the examined penetration rate of AVs. All vehicles are "passenger cars" with dimensions selected randomly, from a default range, by the Aimsun simulator. The desired speed of each vehicle (of both types) is selected randomly, with uniform distribution, from a range [80, 120] km/h. The constant time gap, $\omega$, for both automated and manually driven vehicles, is also selected randomly, with uniform distribution, from a range [0.8, 1.8] s \citep{spiliopoulou2018adaptive}. 

For AVs, a planning horizon of 8 s is used, with a path-planning step of $T=0.25$ s. Thus, each plan consists of controls $j_x (k),a_y (k)$ for $K=32$ time steps. However, as mentioned earlier, half planning horizon is applied at most, which corresponds to 4 s, before a re-plan. The control bounds have been set to $j_x \in [-4.0,4.0]$ and $a_y \in [-1.5,1.5]$ for longitudinal and lateral controls, respectively, but it should be noted that high (absolute) control values are virtually never reached. In an initial offline trial-and-error procedure, where many different driving situations were tested, the penalty weights in \eqref{eq: obj. criterion} were set to $[w_1,w_2,w_3,w_4,w_5,w_6,w_7,w_8]=[1.5,1.0,1.5,0.05,1.0,15.0,15.0,1.0]$. In \eqref{eq: neg-speed} we set  $\epsilon = 0.1$; in \eqref{eq: road-bound} $d=1.4$; and in \eqref{eq:ellipse} $p_1= p_2=18$, which lead to rectangular-like ellipses. 

\subsection{Results}

\begin{figure}
	\centering
		\includegraphics[width=.75\linewidth]{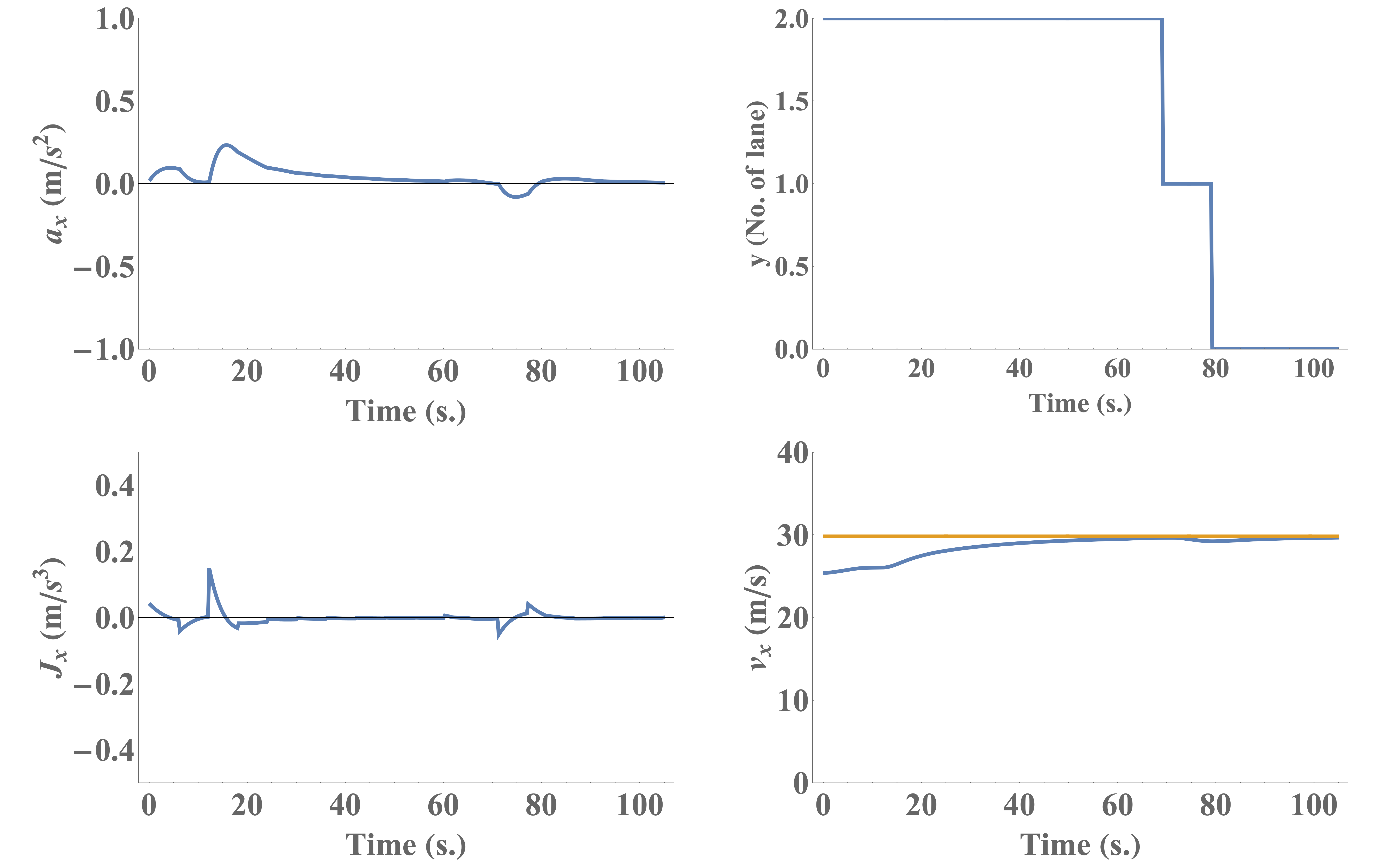}
	\caption{Trajectories of a representative AV during the simulation, representing longitudinal acceleration, lateral position based on lanes, longitudinal jerk and longitudinal speed (orange line is the desired speed).}
	\label{FIG:veh-traj}
\end{figure}

Figure \ref{FIG:veh-traj} displays the trajectories of an AV, extracted from a simulation scenario with low inflow (3.000 veh/h) and 50\% penetration of connected and automated vehicles. The trajectories reflect on the longitudinal acceleration, the lateral position (based on lanes), the longitudinal jerk and speed. The vehicle starts with a low speed and, applying appropriate maneuvers, it manages to reach and maintain a speed close to its desired speed (orange line). Specifically, from the lateral position trajectory, it is noticed that the vehicle applied two lane changes in order to reach its desired speed. The longitudinal acceleration and the corresponding jerk magnitudes are very moderate and smooth, which is good for passenger comfort and fuel consumption.

\begin{figure}[t!]
    \centering
    \begin{subfigure}[t]{0.5\textwidth}
        \centering
        \includegraphics[width=0.95\textwidth]{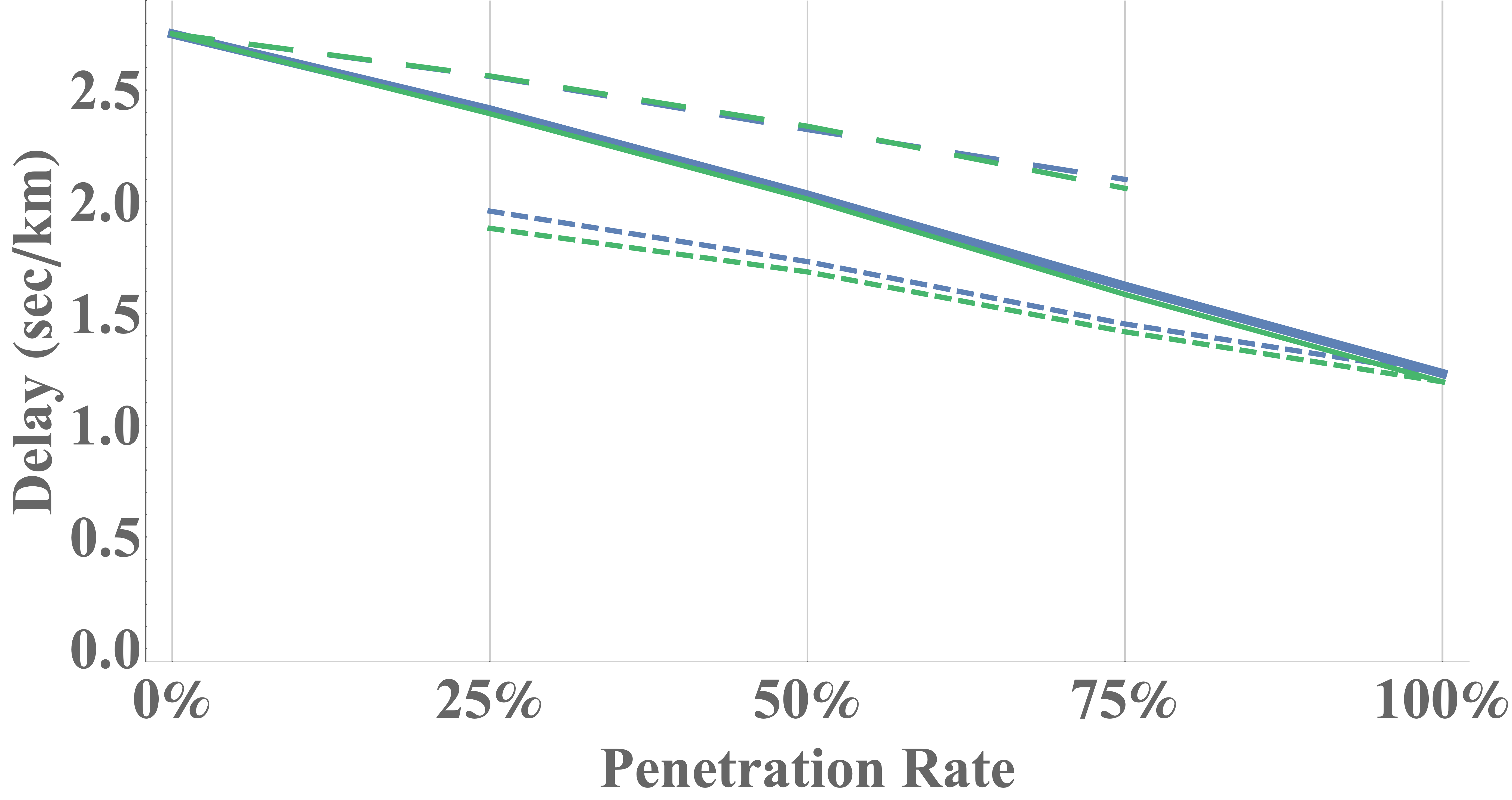}
        \caption{}
        \label{fig:3000-delay}
    \end{subfigure}%
    ~ 
    \begin{subfigure}[t]{0.5\textwidth}
        \centering
        \includegraphics[width=0.95\textwidth]{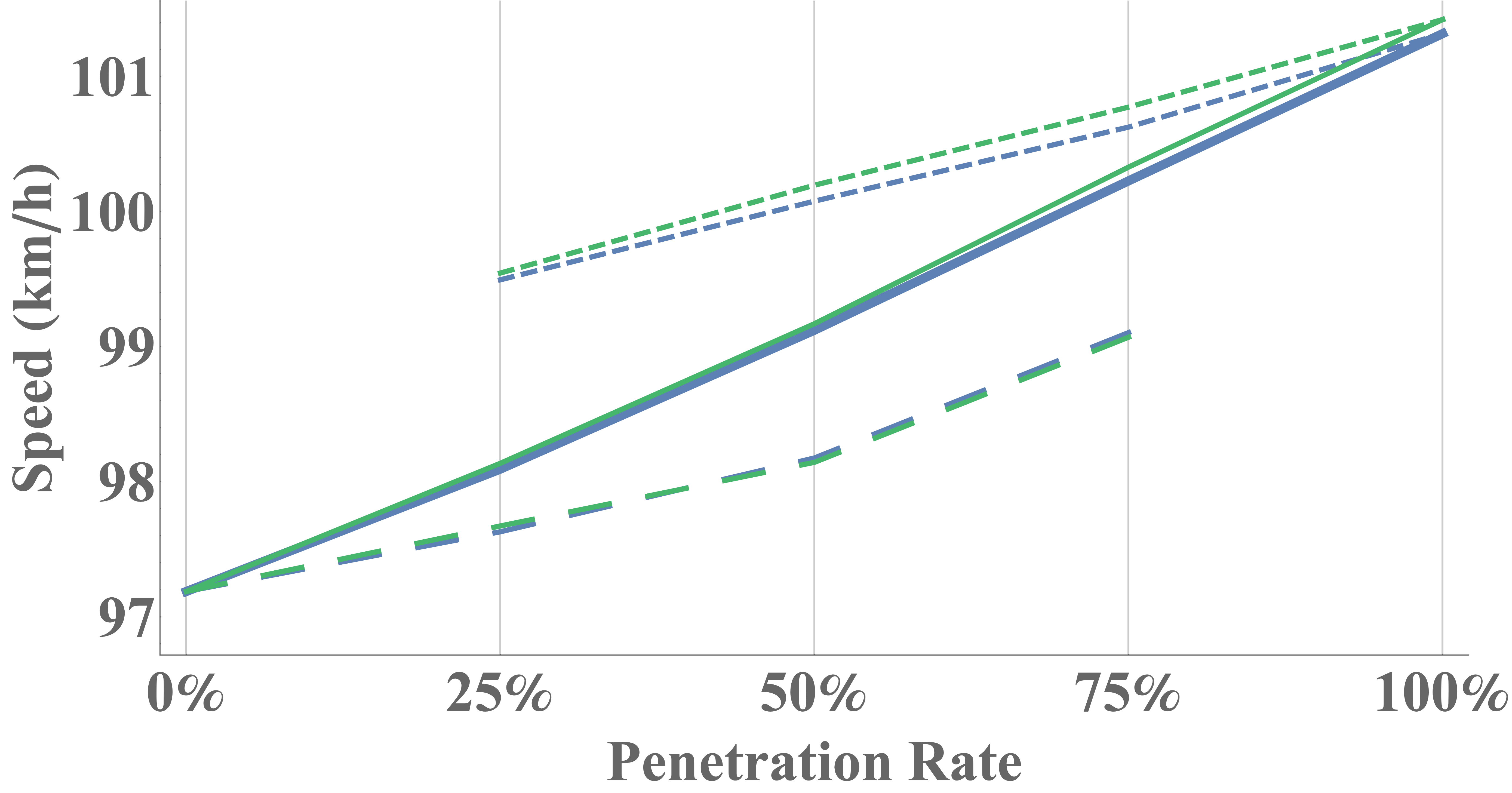}
        \caption{}
        \label{fig:3000-speed}
    \end{subfigure}
    
    \begin{subfigure}[t]{0.5\textwidth}
        \centering
        \includegraphics[width=0.95\textwidth]{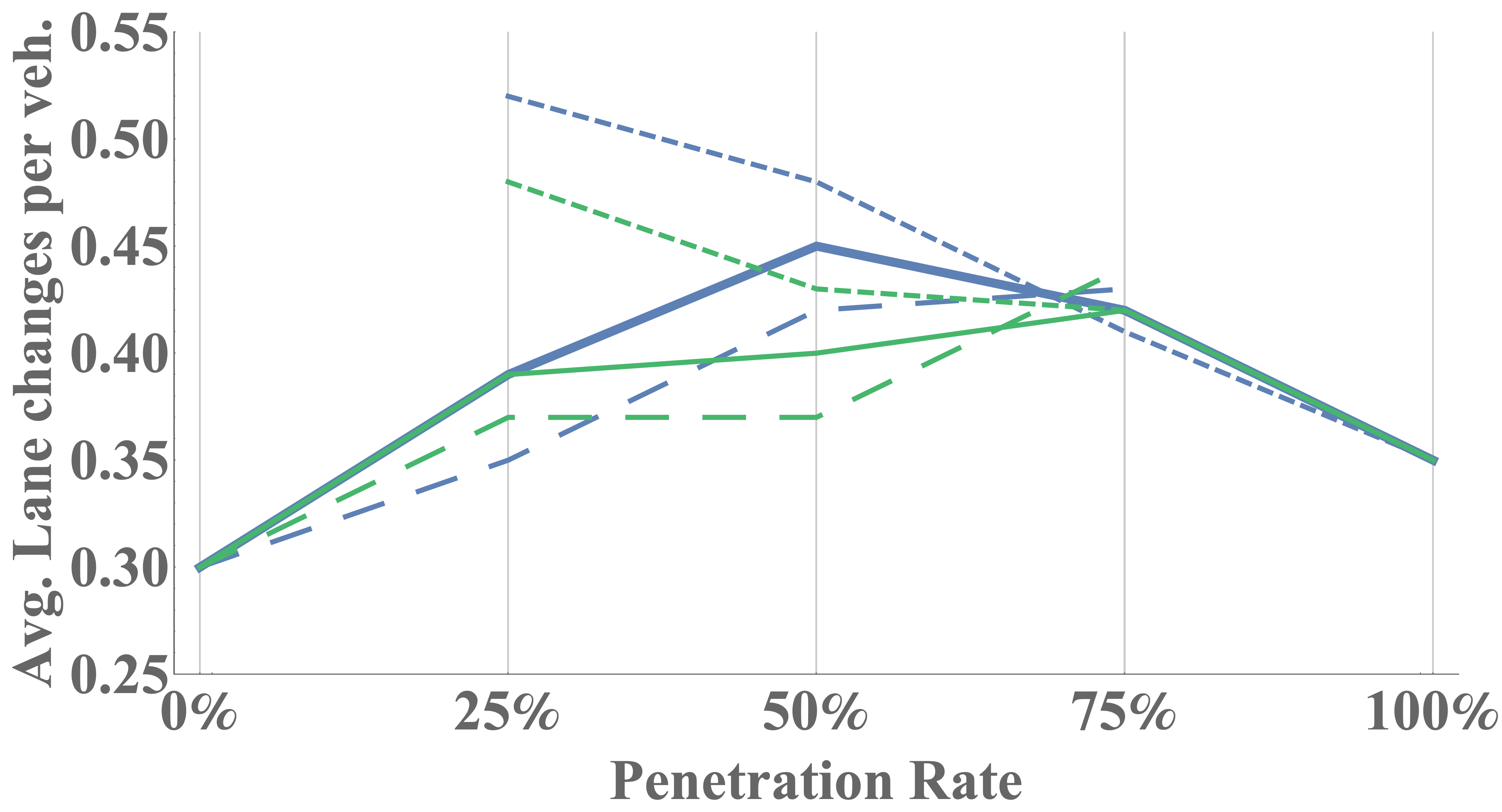}
        \caption{}
        \label{fig:3000-lanes}
    \end{subfigure}%
    ~
    \begin{subfigure}[t]{0.5\textwidth}
        \centering
        \includegraphics[width=0.95\textwidth]{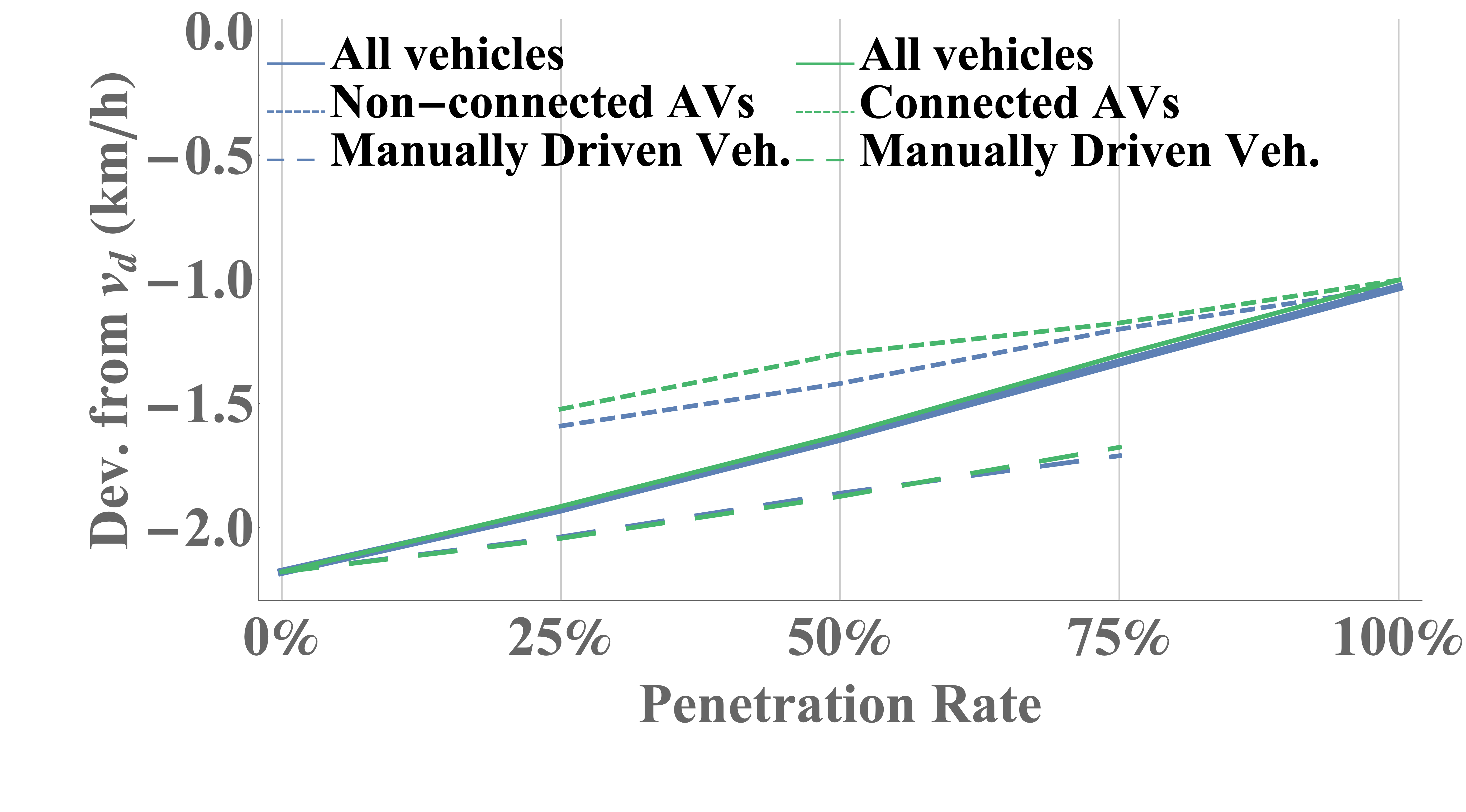}
        \caption{}
        \label{fig:3000-devSpeed}
    \end{subfigure}
    \caption{Average delay time, speed, deviation from target speed and number of lane changes for: i) connected (green lines); and ii) non-connected (blue lines) AVs; for a demand of 3.000 veh/h. Solid lines represent the average of the whole section, while dense-dashed and sparse-dashed lines reflect on average results for automated and manually driven vehicles, respectively.}
    \label{fig: results-3000}
\end{figure}

\begin{figure}[t!]
    \centering
    \begin{subfigure}[t]{0.5\textwidth}
        \centering
        \includegraphics[width=0.95\textwidth]{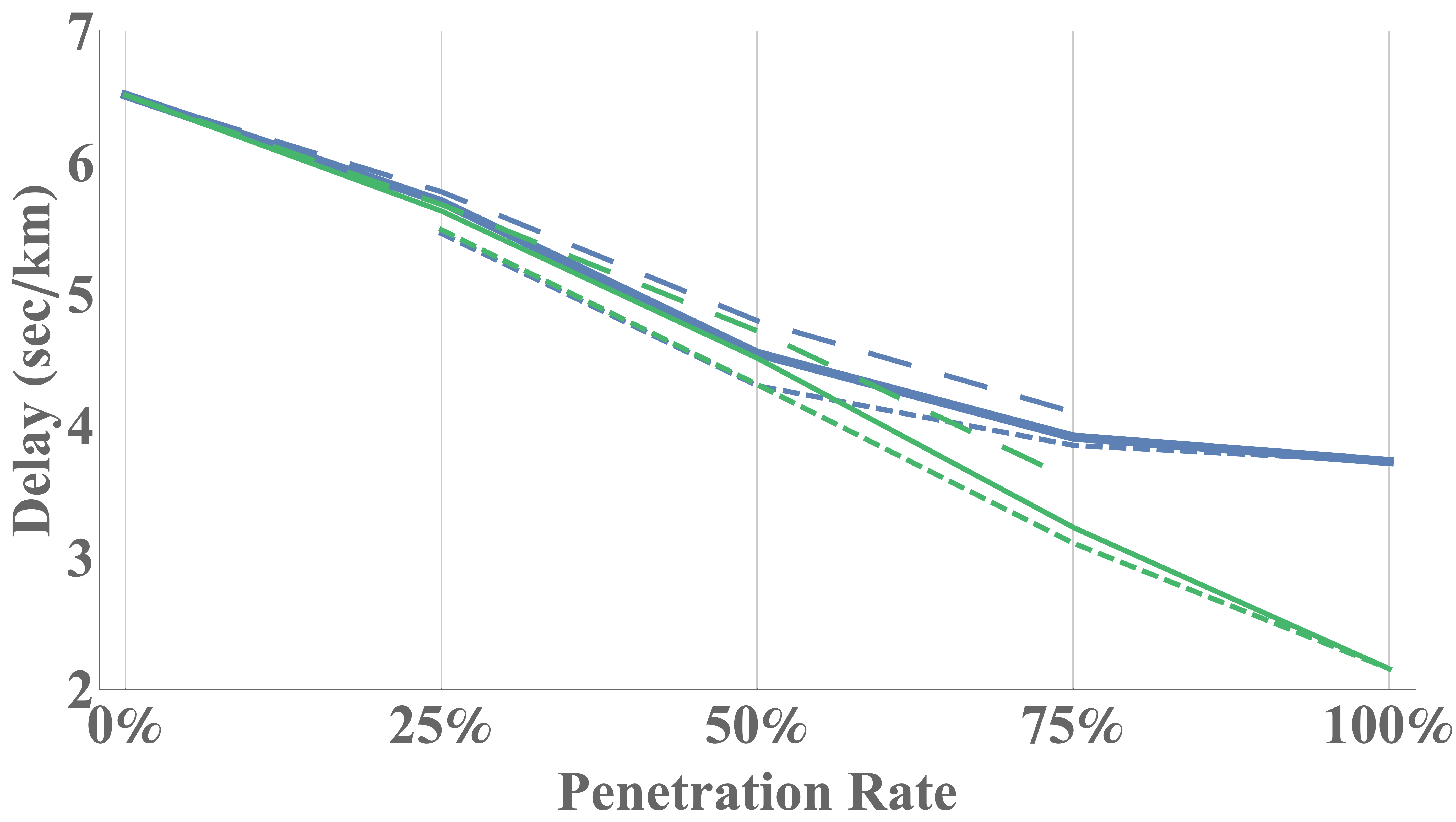}
        \caption{}
        \label{fig:5000-delay}
    \end{subfigure}%
    ~ 
    \begin{subfigure}[t]{0.5\textwidth}
        \centering
        \includegraphics[width=0.95\textwidth]{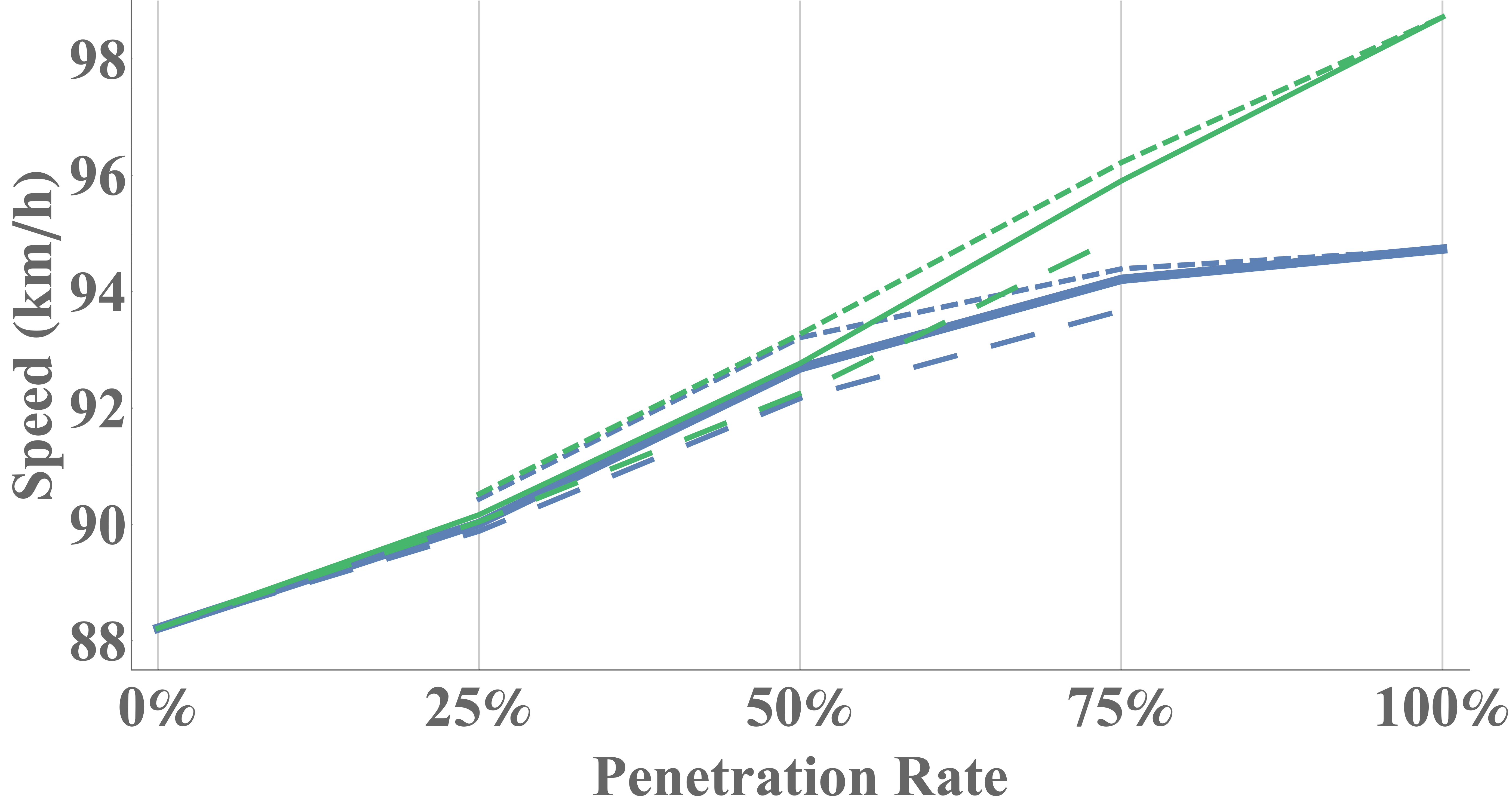}
        \caption{}
        \label{fig:5000-speed}
    \end{subfigure}
    
    \begin{subfigure}[t]{0.5\textwidth}
        \centering
        \includegraphics[width=0.95\textwidth]{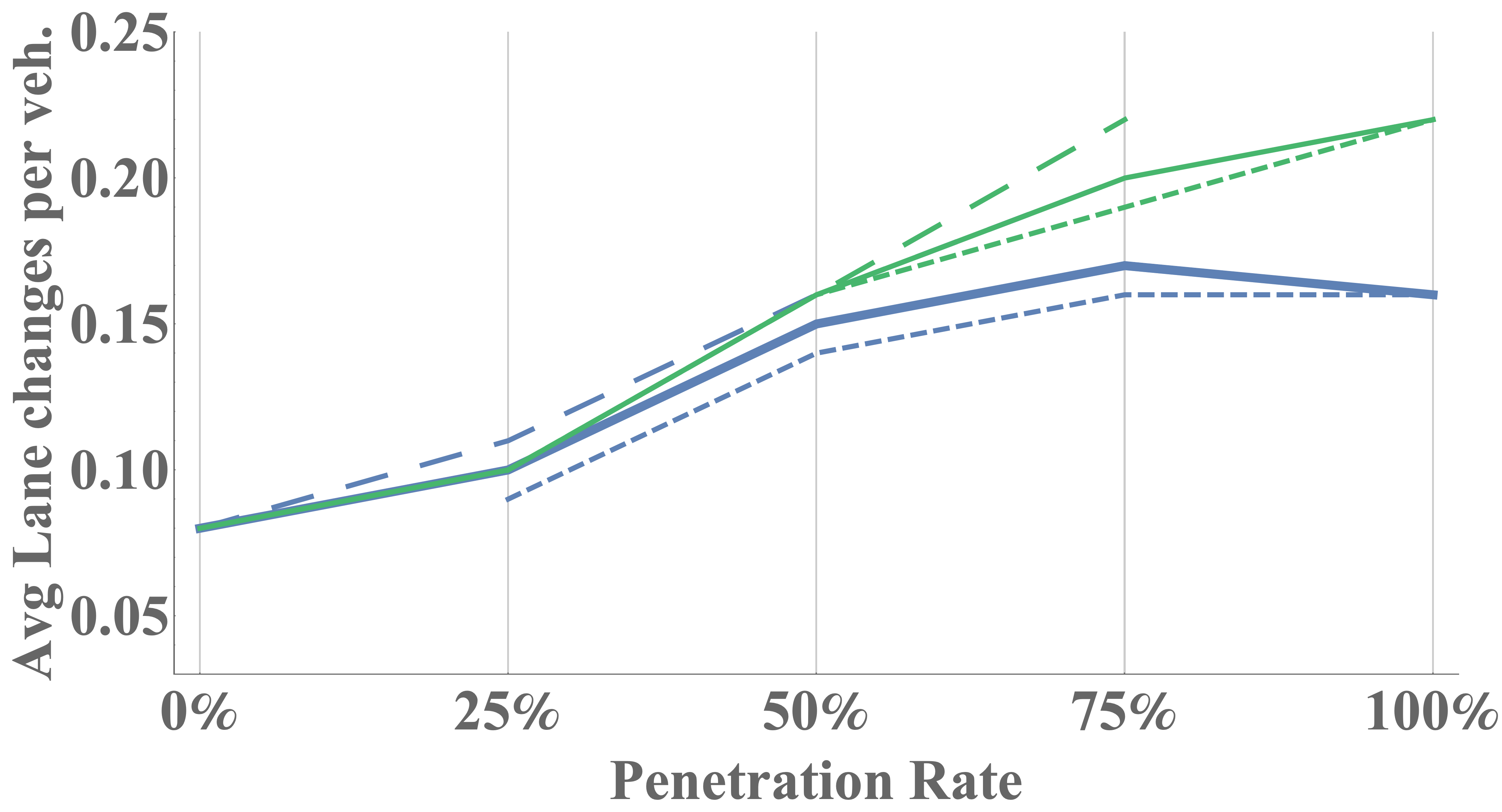}
        \caption{}
        \label{fig:5000-lanes}
    \end{subfigure}%
    ~
    \begin{subfigure}[t]{0.5\textwidth}
        \centering
        \includegraphics[width=0.95\textwidth]{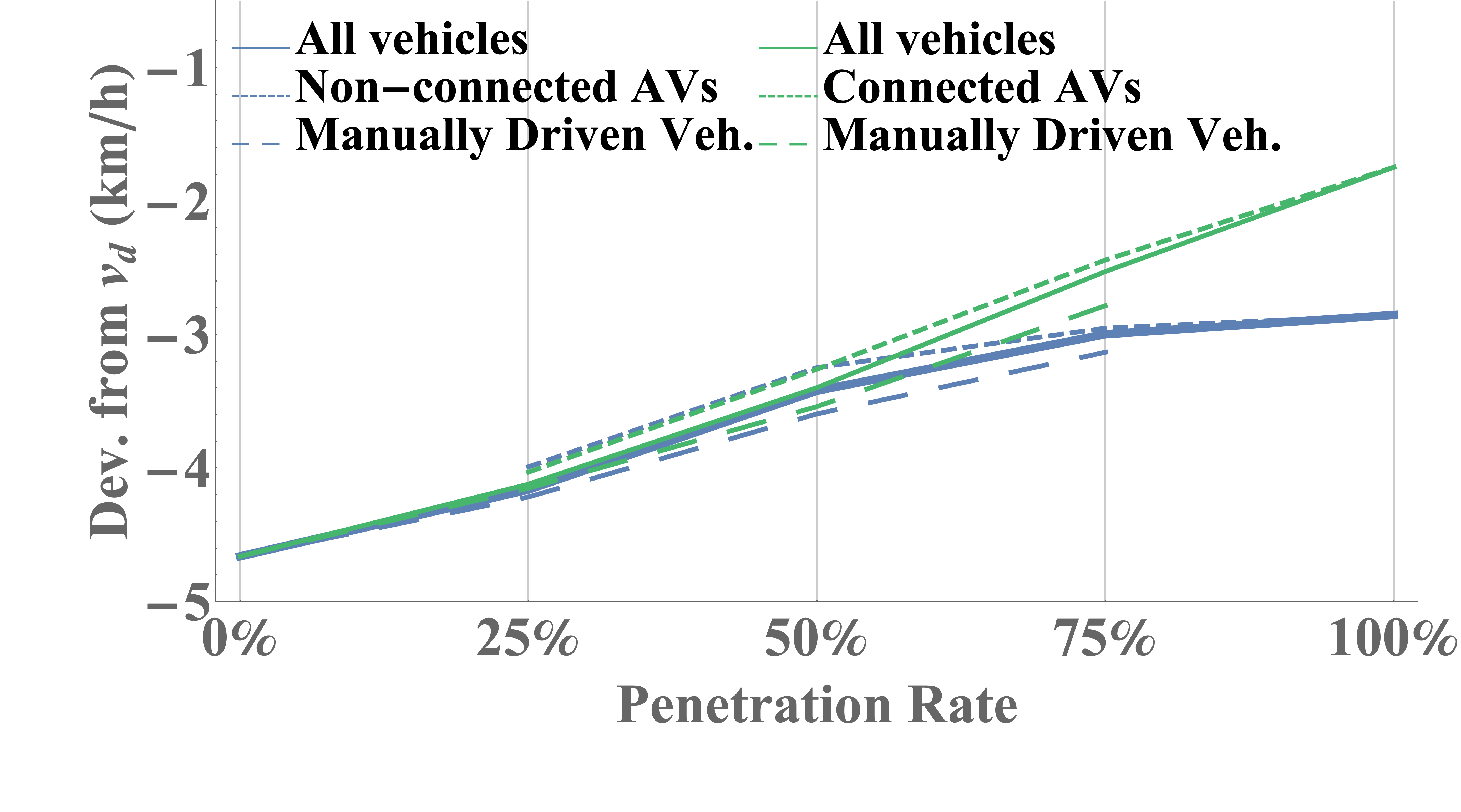}
        \caption{}
        \label{fig:5000-devSpeed}
    \end{subfigure}
    \caption{Average delay time, speed, deviation from target speed and number of lane changes for: i) connected (green lines); and ii) non-connected (blue lines) AVs; for a demand of 5.000 veh/h. Solid lines represent the average of the whole section, while dense-dashed and sparse-dashed lines reflect on average results for automated and manually driven vehicles, respectively.}
    \label{fig: results-5000}
\end{figure}

Figures \ref{fig: results-3000} and \ref{fig: results-5000} present results obtained for the two different demand levels, namely for 3.000 veh/h and 5.000 veh/h, respectively. Each figure contains results of different penetration rates of AVs. In addition, each figure displays and contrasts results corresponding to the two evaluated cases of AV connectivity: connected AVs (green lines) and non-connected AVs (blue lines). For each case, the solid lines reflect on the average results of the whole vehicle population, including both automated and manually driven vehicles, while the dense-dashed and sparse-dashed lines reflect on the average results of automated and manually driven vehicles, respectively. These summarized results concern the average delay time; the average speed; the average number of lane changes; and the average deviation from the desired speed. 

Under both considered demand levels, the online path-planning approach appears to be more efficient in navigating AV speeds closer to the respective desired speeds, compared to manually driven vehicles. Apparently, the suggested approach is more successful at exploiting gaps through traffic and applies "smarter" maneuvers; which leads not only to better performance of each AV, but also to increased overall traffic performance. Specifically, in both Figures \ref{fig: results-3000} and \ref{fig: results-5000}, as the penetration rate of AVs rises, an increase of the average speed, and consequently a significant decrease of the delay, for the whole traffic, is observed. Note that, this improvement also affects the manually driven vehicles, which appear to also benefit from the AVs presence and decisions.

In terms of lane changing, the number of lane changes of all vehicles, in the lower demand level (Figure \ref{fig:3000-lanes}), is higher compared to the higher demand level (Figure \ref{fig:5000-lanes}). This is due to the density prevailing in each case, where, for the lower demand levels there is more space for the vehicles to apply a lane change in order to overtake slower traffic. Moreover, in Figure \ref{fig:3000-lanes}, it can be observed that the number of lane changes of AVs is decreasing as their penetration rate rises. This happens as, for low penetration rates, the AVs need to overtake slower traffic, including other slower AVs or manually driven vehicles, which do not have the same capability to achieve their target speed; while for higher penetration rates, the need for overtaking is reduced, as the increased number of AVs ensures speeds closer to the target speed of each vehicle; thus, it is less probable for a vehicle entering the section to face a slower one. On the other hand, in higher demand levels (Figure \ref{fig: results-5000}), the number of lane changes of AVs is increasing with their increasing penetration. In this case, where traffic is denser, it is harder, for both the manually driven vehicles and the AVs to overtake. Thus, as the penetration rate of AVs rises, there are more overtakes from AVs, which exploit the available space better, helping to maintain increased speed and consequently create spaces for the following traffic. In both demand levels, the AVs efficient maneuvering behavior allows for the manually driven vehicles to increase their lane changes as well.

Contrasting the two types of AVs, the results in Figure \ref{fig: results-3000} indicate only small differences, with connected vehicles being able to achieve slightly better performance in higher penetration rates. This similar outcome is due to the fact that, in lower demand levels, driving space is ample for an AV to maneuver efficiently, which is also evident from Figure \ref{fig:3000-lanes}, where the difference in the average number of lane changes is moderate. For the same reason, all vehicles (manually driven and automated) do not need to change their speed frequently or strongly, hence the added value of receiving improved information (last path decision) from the surrounding AVs, through the connectivity with other vehicles, is not significant. On the other hand, in Figure \ref{fig: results-5000}, where the demand level is higher, vehicle connectivity is seen to have a high impact on the vehicles performance. In this scenario, it is noticed that, although both types of vehicles manage to achieve improved performance, connected AVs outperform the non-connected ones, as the penetration rate increases, with most noticeable differences in the area of 75\%-100\%. This is explained as, in denser conditions, the enhanced information that the AVs have about the surrounding traffic enable them to achieve better reactions in need of a lane change. That means, that the connected AVs are able to apply few lane changes due to better timing, compared to the non-connected ones, which also leads to keeping their speed closer to the desired speed.

\begin{figure}[t!]
    \centering
    \begin{subfigure}[t]{0.5\textwidth}
        \centering
        \includegraphics[height=1.2in]{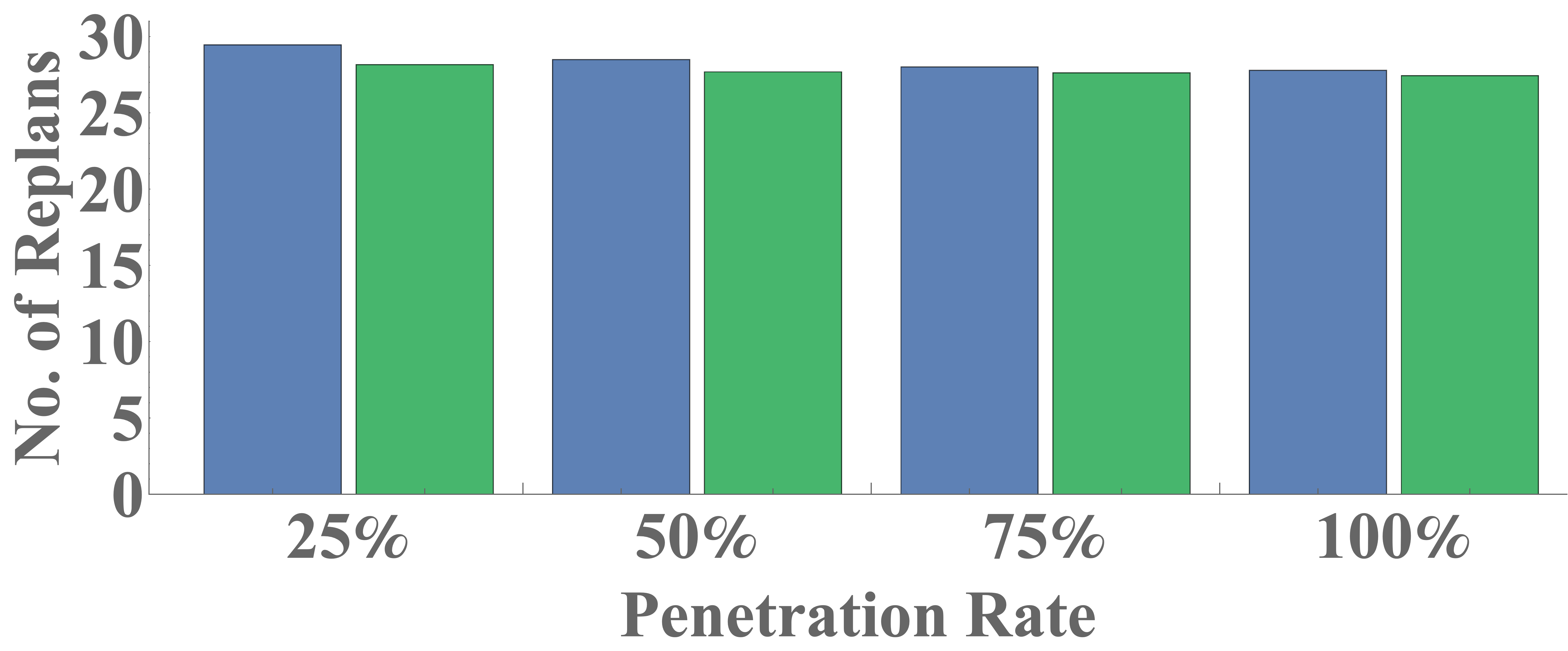}
        \caption{}
        \label{fig: replans-3000}
    \end{subfigure}%
    ~ 
    \begin{subfigure}[t]{0.5\textwidth}
        \centering
        \includegraphics[height=1.2in]{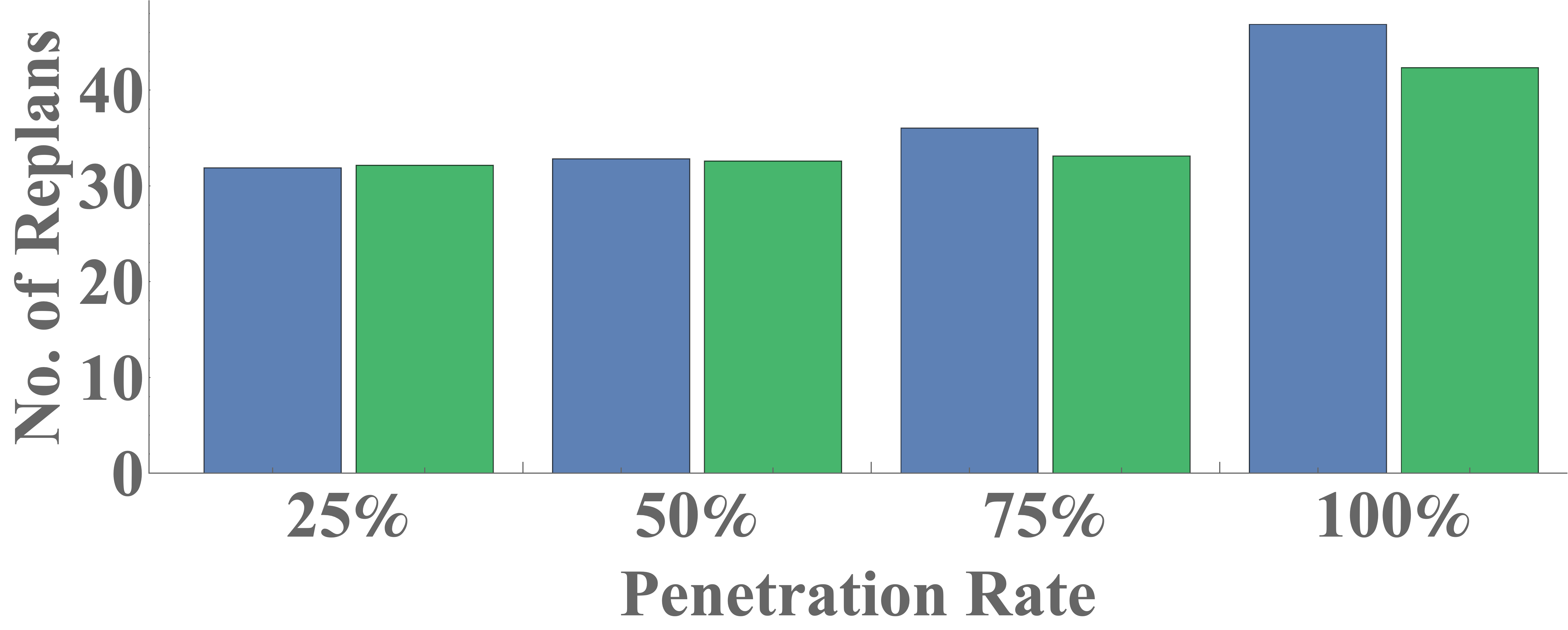}
        \caption{}
         \label{fig: replans-5000}
    \end{subfigure}
    \caption{Average number of (re-)plans for different penetration rates of the two evaluated cases: non-connected AVs (blue); and connected AVs (green) and for different demand levels: a) 3.000 and b) 5.000 veh/h}
    \label{fig: replans}
\end{figure}

Finally, Figure \ref{fig: replans} reports on the average number of plans (and re-plans) of the two types of AVs, in dependence of the penetration rates and for both demand levels. In Figure \ref{fig: replans-3000}, it is noticed that both types of AVs have approximately the same average number of (re-)plans. In this low demand case, all AVs are able to navigate close to their desired speeds, with no significant changes to their predicted paths. The slightly reduced values for the connected AVs are due to their enhanced information, specifically the knowledge of the lane changes that other neighboring AVs have planned to apply. On the other hand, in Figure \ref{fig: replans-5000}, where the higher demand level case is presented, it is observed that, as the penetration rate increases, there is an increase of plan numbers for both types of AVs. As also mentioned above for the lane changing behavior, this is because in this case, for low penetration rates, the surroundings of each AV do not change much, hence there are few deviations from the predicted paths for the obstacles. However, as the penetration rate rises, the AVs are more capable in maneuvering through traffic, which increases the need for re-plans. In these conditions, the number of lane changes increases, vehicles may accelerate or decelerate after an overtake or may reach a vehicle downstream, which was not included in their initial prediction. This increase of the average number of plans is different for the two types of AVs, with connected AVs demanding less re-plans compared to the non-connected ones, as their enhanced knowledge allows them to have more accurate view of the surrounding traffic, including the intended lane changes and the tendency of other AVs to accelerate or decelerate.

Regarding computation times, the average CPU-time per planning of an AV path during the simulation is 0.1 s for the DP, 0.01 s for forward-branching DP and 0.01 s for FDA, with the corresponding maximum values during a whole simulation being 0.2 s for DP, 0.06 s for forward-branching DP and 0.2 s for FDA, which indicates that the proposed approach is clearly real-time feasible.

Demonstration of the vehicle movements in Aimsun micro-simulation platform is available as videos at \url{https://bit.ly/3m2nPa2}. Specifically, there are three videos, showing the connected AVs' behavior in both demand levels, i.e. 3.000 veh/h (video-1) and 5.000 veh/h (video-2 and video-3), both for 100\% penetration rate. It is evident from the videos that the above OCP formulation is conceived for American freeway traffic rules, whereby vehicles may use any lane at any speed and may overtake on the left or right. Adoption of European driving rules, where overtaking is only from left, may be easily accommodated according to \citep{makantasis2018motorway}.

%

\section{Conclusions}
Automated vehicle path-planning has been expressed as an optimal control problem. A combination of DP and NLP techniques allows obtaining good local minima for this non-convex optimization problem efficiently. This efficiency facilitates an online MPC-based (re-)planning approach, by observing and predicting the trajectory of the surrounding vehicles and adapting the EV path accordingly. The proposed MPC-based approach is embedded within the Aimsun micro-simulation platform, enabling us to thoroughly examine the behavior of the approach in presence of vehicles emulating human driving behavior and in a plethora of realistic driving instances. For the simulation investigations, two cases were considered, each of them at different penetration rates of AVs: i) non-connected vehicles, where each AV is aware only of the current states of obstacles, which are either manually driven or automated; and ii) connected AVs, where each AV is aware of the current states of obstacles, but, in addition, it receives (asynchronously) the path-planning decisions of other AVs.

Demonstration results are reported for a homogeneous motorway stretch and different lane-capacity utilizations. Based on the results, it can be seen that the introduction of AVs, guided by the suggested approach, benefits the overall traffic performance. Specifically, in all scenarios, as the penetration of AVs rises, an increase of the average speed and consequently a decrease of the average delay of both automated and manually driven vehicles is noticed. AVs appear to be more effective to navigate closer to the desired speed compared to the manually driven vehicles in all tested scenarios. On the other hand, as far as the AVs are concerned, in lower demand levels both connected and non-connected AVs perform similarly, due to sufficient space for maneuvering. The superiority of connectivity becomes evident in higher demand levels, as the enhanced information about the surrounding traffic is crucial, and connected AVs appear more efficient at increased penetration rates, due to the improved real-time information that enables more pertinent obstacle movement prediction.

\section*{Acknowledgments}
The research leading to these results has received funding form the European Research Council under the European Union’s Horizon 2020 Research and Innovation programme / ERC Grant Agreement n. [833915], project TrafficFluid.

\bibliographystyle{cas-model2-names}

\bibliography{cas-sc-template.bib}

\begin{thebibliography}{35}
\expandafter\ifx\csname natexlab\endcsname\relax\def\natexlab#1{#1}\fi
\providecommand{\url}[1]{\texttt{#1}}
\providecommand{\href}[2]{#2}
\providecommand{\path}[1]{#1}
\providecommand{\DOIprefix}{doi:}
\providecommand{\ArXivprefix}{arXiv:}
\providecommand{\URLprefix}{URL: }
\providecommand{\Pubmedprefix}{pmid:}
\providecommand{\doi}[1]{\href{http://dx.doi.org/#1}{\path{#1}}}
\providecommand{\Pubmed}[1]{\href{pmid:#1}{\path{#1}}}
\providecommand{\bibinfo}[2]{#2}
\ifx\xfnm\relax \def\xfnm[#1]{\unskip,\space#1}\fi
\bibitem[{Aimsun~Next(2019)}]{Aimsun2019}
\bibinfo{author}{Aimsun~Next, .}, \bibinfo{year}{2019}.
\newblock \bibinfo{title}{Transport simulation system (tss)}.
\newblock \bibinfo{note}{\url {https://www.aimsun.com/aimsun-next}}.
\bibitem[{Claussmann et~al.(2019)Claussmann, Revilloud, Gruyer and
  Glaser}]{claussmann2019review}
\bibinfo{author}{Claussmann, L.}, \bibinfo{author}{Revilloud, M.},
  \bibinfo{author}{Gruyer, D.}, \bibinfo{author}{Glaser, S.},
  \bibinfo{year}{2019}.
\newblock \bibinfo{title}{A review of motion planning for highway autonomous
  driving}.
\newblock \bibinfo{journal}{IEEE Transactions on Intelligent Transportation
  Systems} \bibinfo{volume}{21}, \bibinfo{pages}{1826--1848}.
\bibitem[{Dey et~al.(2015)Dey, Yan, Wang, Wang, Shen, Chowdhury, Yu, Qiu and
  Soundararaj}]{dey2015review}
\bibinfo{author}{Dey, K.C.}, \bibinfo{author}{Yan, L.}, \bibinfo{author}{Wang,
  X.}, \bibinfo{author}{Wang, Y.}, \bibinfo{author}{Shen, H.},
  \bibinfo{author}{Chowdhury, M.}, \bibinfo{author}{Yu, L.},
  \bibinfo{author}{Qiu, C.}, \bibinfo{author}{Soundararaj, V.},
  \bibinfo{year}{2015}.
\newblock \bibinfo{title}{A review of communication, driver characteristics,
  and controls aspects of cooperative adaptive cruise control (cacc)}.
\newblock \bibinfo{journal}{IEEE Transactions on Intelligent Transportation
  Systems} \bibinfo{volume}{17}, \bibinfo{pages}{491--509}.
\bibitem[{Dixit et~al.(2019)Dixit, Montanaro, Dianati, Oxtoby, Mizutani,
  Mouzakitis and Fallah}]{dixit2019trajectory}
\bibinfo{author}{Dixit, S.}, \bibinfo{author}{Montanaro, U.},
  \bibinfo{author}{Dianati, M.}, \bibinfo{author}{Oxtoby, D.},
  \bibinfo{author}{Mizutani, T.}, \bibinfo{author}{Mouzakitis, A.},
  \bibinfo{author}{Fallah, S.}, \bibinfo{year}{2019}.
\newblock \bibinfo{title}{Trajectory planning for autonomous high-speed
  overtaking in structured environments using robust mpc}.
\newblock \bibinfo{journal}{IEEE Transactions on Intelligent Transportation
  Systems} \bibinfo{volume}{21}, \bibinfo{pages}{2310--2323}.
\bibitem[{Earl and D’Andrea(2007)}]{earl2007decomposition}
\bibinfo{author}{Earl, M.G.}, \bibinfo{author}{D’Andrea, R.},
  \bibinfo{year}{2007}.
\newblock \bibinfo{title}{A decomposition approach to multi-vehicle cooperative
  control}.
\newblock \bibinfo{journal}{Robotics and Autonomous Systems}
  \bibinfo{volume}{55}, \bibinfo{pages}{276--291}.
\bibitem[{Ghiasi et~al.(2017)Ghiasi, Ma, Zhou and Li}]{ghiasi2017speed}
\bibinfo{author}{Ghiasi, A.}, \bibinfo{author}{Ma, J.}, \bibinfo{author}{Zhou,
  F.}, \bibinfo{author}{Li, X.}, \bibinfo{year}{2017}.
\newblock \bibinfo{title}{Speed harmonization algorithm using connected
  autonomous vehicles}, in: \bibinfo{booktitle}{96th Annual Meeting of the
  Transportation Research Board}.
\bibitem[{Gipps(1986)}]{gipps1986model}
\bibinfo{author}{Gipps, P.G.}, \bibinfo{year}{1986}.
\newblock \bibinfo{title}{A model for the structure of lane-changing
  decisions}.
\newblock \bibinfo{journal}{Transportation Research Part B: Methodological}
  \bibinfo{volume}{20}, \bibinfo{pages}{403--414}.
\bibitem[{Goniewicz et~al.(2016)Goniewicz, Goniewicz, Paw{\l}owski and
  Fiedor}]{goniewicz2016road}
\bibinfo{author}{Goniewicz, K.}, \bibinfo{author}{Goniewicz, M.},
  \bibinfo{author}{Paw{\l}owski, W.}, \bibinfo{author}{Fiedor, P.},
  \bibinfo{year}{2016}.
\newblock \bibinfo{title}{Road accident rates: strategies and programmes for
  improving road traffic safety}.
\newblock \bibinfo{journal}{European journal of trauma and emergency surgery}
  \bibinfo{volume}{42}, \bibinfo{pages}{433--438}.
\bibitem[{Gonz{\'a}lez et~al.(2015)Gonz{\'a}lez, P{\'e}rez, Milan{\'e}s and
  Nashashibi}]{gonzalez2015review}
\bibinfo{author}{Gonz{\'a}lez, D.}, \bibinfo{author}{P{\'e}rez, J.},
  \bibinfo{author}{Milan{\'e}s, V.}, \bibinfo{author}{Nashashibi, F.},
  \bibinfo{year}{2015}.
\newblock \bibinfo{title}{A review of motion planning techniques for automated
  vehicles}.
\newblock \bibinfo{journal}{IEEE Transactions on Intelligent Transportation
  Systems} \bibinfo{volume}{17}, \bibinfo{pages}{1135--1145}.
\bibitem[{Gu and Hu(2002)}]{gu2002neural}
\bibinfo{author}{Gu, D.}, \bibinfo{author}{Hu, H.}, \bibinfo{year}{2002}.
\newblock \bibinfo{title}{Neural predictive control for a car-like mobile
  robot}.
\newblock \bibinfo{journal}{Robotics and Autonomous Systems}
  \bibinfo{volume}{39}, \bibinfo{pages}{73--86}.
\bibitem[{Gu and Dolan(2014)}]{gu2014toward}
\bibinfo{author}{Gu, T.}, \bibinfo{author}{Dolan, J.M.}, \bibinfo{year}{2014}.
\newblock \bibinfo{title}{Toward human-like motion planning in urban
  environments}, in: \bibinfo{booktitle}{2014 IEEE Intelligent Vehicles
  Symposium Proceedings}, \bibinfo{organization}{IEEE}. pp.
  \bibinfo{pages}{350--355}.
\bibitem[{{Haydari} and {Yilmaz}(2020)}]{haydari2020survey}
\bibinfo{author}{{Haydari}, A.}, \bibinfo{author}{{Yilmaz}, Y.},
  \bibinfo{year}{2020}.
\newblock \bibinfo{title}{Deep reinforcement learning for intelligent
  transportation systems: A survey}.
\newblock \bibinfo{journal}{IEEE Transactions on Intelligent Transportation
  Systems} , \bibinfo{pages}{1--22}\DOIprefix\doi{10.1109/TITS.2020.3008612}.
\bibitem[{Jalalmaab et~al.(2015)Jalalmaab, Fidan, Jeon and
  Falcone}]{jalalmaab2015model}
\bibinfo{author}{Jalalmaab, M.}, \bibinfo{author}{Fidan, B.},
  \bibinfo{author}{Jeon, S.}, \bibinfo{author}{Falcone, P.},
  \bibinfo{year}{2015}.
\newblock \bibinfo{title}{Model predictive path planning with time-varying
  safety constraints for highway autonomous driving}, in:
  \bibinfo{booktitle}{2015 International Conference on Advanced Robotics
  (ICAR)}, \bibinfo{organization}{IEEE}. pp. \bibinfo{pages}{213--217}.
\bibitem[{Khazaeni and Cassandras(2016)}]{khazaeni2016event}
\bibinfo{author}{Khazaeni, Y.}, \bibinfo{author}{Cassandras, C.G.},
  \bibinfo{year}{2016}.
\newblock \bibinfo{title}{Event-driven cooperative receding horizon control for
  multi-agent systems in uncertain environments}.
\newblock \bibinfo{journal}{IEEE Transactions on Control of Network Systems}
  \bibinfo{volume}{5}, \bibinfo{pages}{409--422}.
\bibitem[{Makantasis and Papageorgiou(2018)}]{makantasis2018motorway}
\bibinfo{author}{Makantasis, K.}, \bibinfo{author}{Papageorgiou, M.},
  \bibinfo{year}{2018}.
\newblock \bibinfo{title}{Motorway path planning for automated road vehicles
  based on optimal control methods}.
\newblock \bibinfo{journal}{Transportation Research Record}
  \bibinfo{volume}{2672}, \bibinfo{pages}{112--123}.
\bibitem[{Malikopoulos et~al.(2018)Malikopoulos, Hong, Park, Lee and
  Ryu}]{malikopoulos2018optimal}
\bibinfo{author}{Malikopoulos, A.A.}, \bibinfo{author}{Hong, S.},
  \bibinfo{author}{Park, B.B.}, \bibinfo{author}{Lee, J.},
  \bibinfo{author}{Ryu, S.}, \bibinfo{year}{2018}.
\newblock \bibinfo{title}{Optimal control for speed harmonization of automated
  vehicles}.
\newblock \bibinfo{journal}{IEEE Transactions on Intelligent Transportation
  Systems} \bibinfo{volume}{20}, \bibinfo{pages}{2405--2417}.
\bibitem[{Mayne(2014)}]{mayne2014model}
\bibinfo{author}{Mayne, D.Q.}, \bibinfo{year}{2014}.
\newblock \bibinfo{title}{Model predictive control: Recent developments and
  future promise}.
\newblock \bibinfo{journal}{Automatica} \bibinfo{volume}{50},
  \bibinfo{pages}{2967--2986}.
\bibitem[{Mayne and Michalska(1988)}]{mayne1988receding}
\bibinfo{author}{Mayne, D.Q.}, \bibinfo{author}{Michalska, H.},
  \bibinfo{year}{1988}.
\newblock \bibinfo{title}{Receding horizon control of nonlinear systems}, in:
  \bibinfo{booktitle}{Proceedings of the 27th IEEE Conference on Decision and
  Control}, \bibinfo{organization}{IEEE}. pp. \bibinfo{pages}{464--465}.
\bibitem[{Montanaro et~al.(2019)Montanaro, Dixit, Fallah, Dianati, Stevens,
  Oxtoby and Mouzakitis}]{montanaro2019towards}
\bibinfo{author}{Montanaro, U.}, \bibinfo{author}{Dixit, S.},
  \bibinfo{author}{Fallah, S.}, \bibinfo{author}{Dianati, M.},
  \bibinfo{author}{Stevens, A.}, \bibinfo{author}{Oxtoby, D.},
  \bibinfo{author}{Mouzakitis, A.}, \bibinfo{year}{2019}.
\newblock \bibinfo{title}{Towards connected autonomous driving: review of
  use-cases}.
\newblock \bibinfo{journal}{Vehicle system dynamics} \bibinfo{volume}{57},
  \bibinfo{pages}{779--814}.
\bibitem[{Nagy and Kelly(2001)}]{nagy2001trajectory}
\bibinfo{author}{Nagy, B.}, \bibinfo{author}{Kelly, A.}, \bibinfo{year}{2001}.
\newblock \bibinfo{title}{Trajectory generation for car-like robots using cubic
  curvature polynomials}.
\newblock \bibinfo{journal}{Field and Service Robots} \bibinfo{volume}{11}.
\bibitem[{Ntousakis et~al.(2016)Ntousakis, Nikolos and
  Papageorgiou}]{ntousakis2016optimal}
\bibinfo{author}{Ntousakis, I.A.}, \bibinfo{author}{Nikolos, I.K.},
  \bibinfo{author}{Papageorgiou, M.}, \bibinfo{year}{2016}.
\newblock \bibinfo{title}{Optimal vehicle trajectory planning in the context of
  cooperative merging on highways}.
\newblock \bibinfo{journal}{Transportation research part C: emerging
  technologies} \bibinfo{volume}{71}, \bibinfo{pages}{464--488}.
\bibitem[{Papageorgiou et~al.(2015)Papageorgiou, Leibold and
  Buss}]{papageorgiou2015optimierung}
\bibinfo{author}{Papageorgiou, M.}, \bibinfo{author}{Leibold, M.},
  \bibinfo{author}{Buss, M.}, \bibinfo{year}{2015}.
\newblock \bibinfo{title}{Optimierung}. volume~\bibinfo{volume}{4}.
\newblock \bibinfo{publisher}{Springer}.
\bibitem[{Papageorgiou et~al.(2016)Papageorgiou, Marinaki, Typaldos and
  Makantasis}]{papageorgiou2016feasible}
\bibinfo{author}{Papageorgiou, M.}, \bibinfo{author}{Marinaki, M.},
  \bibinfo{author}{Typaldos, P.}, \bibinfo{author}{Makantasis, K.},
  \bibinfo{year}{2016}.
\newblock \bibinfo{title}{A feasible direction algorithm for the numerical
  solution of optimal control problems--extended version}.
\newblock \bibinfo{journal}{Chania, Greece: Technical University of Crete,
  Dynamics Sysyems and Simulations Laboratory} , \bibinfo{pages}{2016--26}.
\bibitem[{Qin and Badgwell(2003)}]{qin2003survey}
\bibinfo{author}{Qin, S.J.}, \bibinfo{author}{Badgwell, T.A.},
  \bibinfo{year}{2003}.
\newblock \bibinfo{title}{A survey of industrial model predictive control
  technology}.
\newblock \bibinfo{journal}{Control engineering practice} \bibinfo{volume}{11},
  \bibinfo{pages}{733--764}.
\bibitem[{Rajamani(2011)}]{rajamani2011vehicle}
\bibinfo{author}{Rajamani, R.}, \bibinfo{year}{2011}.
\newblock \bibinfo{title}{Vehicle dynamics and control}.
\newblock \bibinfo{publisher}{Springer Science \& Business Media}.
\bibitem[{Rasekhipour et~al.(2016)Rasekhipour, Khajepour, Chen and
  Litkouhi}]{rasekhipour2016potential}
\bibinfo{author}{Rasekhipour, Y.}, \bibinfo{author}{Khajepour, A.},
  \bibinfo{author}{Chen, S.K.}, \bibinfo{author}{Litkouhi, B.},
  \bibinfo{year}{2016}.
\newblock \bibinfo{title}{A potential field-based model predictive
  path-planning controller for autonomous road vehicles}.
\newblock \bibinfo{journal}{IEEE Transactions on Intelligent Transportation
  Systems} \bibinfo{volume}{18}, \bibinfo{pages}{1255--1267}.
\bibitem[{Rios-Torres and Malikopoulos(2016)}]{rios2016survey}
\bibinfo{author}{Rios-Torres, J.}, \bibinfo{author}{Malikopoulos, A.A.},
  \bibinfo{year}{2016}.
\newblock \bibinfo{title}{A survey on the coordination of connected and
  automated vehicles at intersections and merging at highway on-ramps}.
\newblock \bibinfo{journal}{IEEE Transactions on Intelligent Transportation
  Systems} \bibinfo{volume}{18}, \bibinfo{pages}{1066--1077}.
\bibitem[{Sjoberg et~al.(2017)Sjoberg, Andres, Buburuzan and
  Brakemeier}]{sjoberg2017cooperative}
\bibinfo{author}{Sjoberg, K.}, \bibinfo{author}{Andres, P.},
  \bibinfo{author}{Buburuzan, T.}, \bibinfo{author}{Brakemeier, A.},
  \bibinfo{year}{2017}.
\newblock \bibinfo{title}{Cooperative intelligent transport systems in europe:
  Current deployment status and outlook}.
\newblock \bibinfo{journal}{IEEE Vehicular Technology Magazine}
  \bibinfo{volume}{12}, \bibinfo{pages}{89--97}.
\bibitem[{Spiliopoulou et~al.(2018)Spiliopoulou, Manolis, Vandorou and
  Papageorgiou}]{spiliopoulou2018adaptive}
\bibinfo{author}{Spiliopoulou, A.}, \bibinfo{author}{Manolis, D.},
  \bibinfo{author}{Vandorou, F.}, \bibinfo{author}{Papageorgiou, M.},
  \bibinfo{year}{2018}.
\newblock \bibinfo{title}{Adaptive cruise control operation for improved
  motorway traffic flow}.
\newblock \bibinfo{journal}{Transportation research record}
  \bibinfo{volume}{2672}, \bibinfo{pages}{24--35}.
\bibitem[{Tian et~al.(2018)Tian, Wu, Boriboonsomsin and
  Barth}]{tian2018performance}
\bibinfo{author}{Tian, D.}, \bibinfo{author}{Wu, G.},
  \bibinfo{author}{Boriboonsomsin, K.}, \bibinfo{author}{Barth, M.J.},
  \bibinfo{year}{2018}.
\newblock \bibinfo{title}{Performance measurement evaluation framework and
  co-benefit$\backslash$/tradeoff analysis for connected and automated vehicles
  (cav) applications: A survey}.
\newblock \bibinfo{journal}{IEEE Intelligent Transportation Systems Magazine}
  \bibinfo{volume}{10}, \bibinfo{pages}{110--122}.
\bibitem[{Treiber and Kesting(2013)}]{treiber2013traffic}
\bibinfo{author}{Treiber, M.}, \bibinfo{author}{Kesting, A.},
  \bibinfo{year}{2013}.
\newblock \bibinfo{title}{Traffic flow dynamics}.
\newblock \bibinfo{journal}{Traffic Flow Dynamics: Data, Models and Simulation,
  Springer-Verlag Berlin Heidelberg} .
\bibitem[{Typaldos et~al.(2020)Typaldos, Papamichail and
  Papageorgiou}]{typaldos2020minimization}
\bibinfo{author}{Typaldos, P.}, \bibinfo{author}{Papamichail, I.},
  \bibinfo{author}{Papageorgiou, M.}, \bibinfo{year}{2020}.
\newblock \bibinfo{title}{Minimization of fuel consumption for vehicle
  trajectories}.
\newblock \bibinfo{journal}{IEEE Transactions on Intelligent Transportation
  Systems} \bibinfo{volume}{21}, \bibinfo{pages}{1716--1727}.
\bibitem[{Wang et~al.(2018a)Wang, Huang, Khajepour, Liu, Qin and
  Zhang}]{wang2018local}
\bibinfo{author}{Wang, H.}, \bibinfo{author}{Huang, Y.},
  \bibinfo{author}{Khajepour, A.}, \bibinfo{author}{Liu, T.},
  \bibinfo{author}{Qin, Y.}, \bibinfo{author}{Zhang, Y.},
  \bibinfo{year}{2018}a.
\newblock \bibinfo{title}{Local path planning for autonomous vehicles: Crash
  mitigation}, in: \bibinfo{booktitle}{2018 IEEE Intelligent Vehicles Symposium
  (IV)}, \bibinfo{organization}{IEEE}. pp. \bibinfo{pages}{1602--1606}.
\bibitem[{Wang et~al.(2018b)Wang, Li and Yao}]{wang2018review}
\bibinfo{author}{Wang, Y.}, \bibinfo{author}{Li, X.}, \bibinfo{author}{Yao,
  H.}, \bibinfo{year}{2018}b.
\newblock \bibinfo{title}{Review of trajectory optimisation for connected
  automated vehicles}.
\newblock \bibinfo{journal}{IET Intelligent Transport Systems}
  \bibinfo{volume}{13}, \bibinfo{pages}{580--586}.
\bibitem[{Wang et~al.(2018c)Wang, Wu and Barth}]{wan2018review}
\bibinfo{author}{Wang, Z.}, \bibinfo{author}{Wu, G.}, \bibinfo{author}{Barth,
  M.J.}, \bibinfo{year}{2018}c.
\newblock \bibinfo{title}{A review on cooperative adaptive cruise control
  (cacc) systems: Architectures, controls, and applications}, in:
  \bibinfo{booktitle}{2018 21st International Conference on Intelligent
  Transportation Systems (ITSC)}, \bibinfo{organization}{IEEE}. pp.
  \bibinfo{pages}{2884--2891}.

\end{thebibliography}

\end{document}